\documentclass[manuscript,screen]{acmart}
\usepackage{xcolor}

\newcommand{\eg}{\emph{e.g., }}

\newlength\myindent
\newtheorem{theorem}{Theorem}[section]

\floatname{algorithm}{Algorithm}

\usepackage{hyperref}
\usepackage{amsmath}
\usepackage{graphicx}
\usepackage{multirow}
\usepackage{adjustbox}
\usepackage{tabularx}
\usepackage{booktabs}
\usepackage{tablefootnote}
\usepackage{booktabs}
\usepackage{array}
\usepackage{balance} 
\usepackage{multirow}
\usepackage[normalem]{ulem}
\usepackage{color}
\definecolor{lightgray}{RGB}{215,215,215}
\usepackage{colortbl}  
\usepackage{xcolor}
\useunder{\uline}{\ul}{}
\usepackage{subfigure}
\usepackage{algorithm}  
\usepackage{algorithmicx}  
\usepackage[noend]{algpseudocode}  

\usepackage{amsmath}  
\usepackage[inline]{enumitem}
\usepackage{tabularx}
\usepackage[utf8]{inputenc}
\usepackage[english]{babel}
\usepackage{amsthm}
\usepackage{bm}
\usepackage{acronym}
\usepackage{mdframed} 
\usepackage{lipsum}   
\acrodef{LLM}{large language model}
\acrodef{CF}{collaborative filtering}
\author{Jujia Zhao}
\email{j.zhao@liacs.leidenuniv.nl}
\affiliation{
\institution{Leiden University}
\country{The Netherlands}
}
\author{Wenjie Wang}
\email{wenjiewang96@gmail.com}
\affiliation{
\institution{University of Science and Technology of China}
\country{China}
}
\author{Chen Xu}
\email{xc_chen@ruc.edu.cn}
\affiliation{
\institution{Renmin University of China}
\country{China}
}
\author{Xiuying Chen}
\email{xiuying.chen@mbzuai.ac.ae}
\affiliation{
\institution{Mohamed bin Zayed University of Artificial Intelligence}
\country{United Arab Emirates}
}
\author{Zhaochun Ren}
\email{z.ren@liacs.leidenuniv.nl}
\affiliation{
\institution{Leiden University}
\country{The Netherlands}
}
\author{Suzan Verberne}
\email{s.verberne@liacs.leidenuniv.nl}
\affiliation{
\institution{Leiden University}
\country{The Netherlands}
}

\begin{document}



\title{Unifying Search and Recommendation with Dual-View Representation Learning in a Generative Paradigm}

\begin{abstract}
Recommender systems and search engines are critical components of online platforms, with recommender systems proactively providing information to users, whereas search engines enable users to seek information actively.
Unifying both tasks in a shared model is promising since it can enhance user modeling and item understanding.
Previous approaches mainly follow a discriminative paradigm, utilizing shared encoders to process input features and task-specific heads to perform each task. 
However, this paradigm faces two key challenges: 
gradient conflict and manual architecture design overhead.
From the information theory perspective, these challenges potentially both stem from the same issue --- low mutual information between input features and task-specific output during optimization.


To address this, we propose GenSR, a novel generative paradigm for unifying search and recommendation (S\&R), which leverages task-specific prompts to partition the model’s parameter space into subspaces, thereby enhancing mutual information.
To construct effective subspaces for each task, GenSR first prepares informative representations for each subspace and then optimizes both subspaces in one unified model.
Specifically, GenSR consists of two main modules:
\begin{enumerate*}[label=(\arabic*)]
\item Dual Representation Learning, which independently models collaborative and semantic historical information to derive expressive item representations; and 
\item S\&R Task Unifying, which utilizes contrastive learning together with instruction tuning to generate task-specific outputs effectively.
\end{enumerate*}
Extensive experiments on two benchmarks show GenSR outperforms state-of-the-art methods across S\&R tasks.


\end{abstract}



\maketitle

\section{Introduction}
\label{sec:introduction}

Recommender systems and search engines have become indispensable components of modern online service platforms, including e-commerce websites and social media networks~\cite{zhang2024towards,wu2024generative}.
Usually, search and recommendation (S\&R) tasks are trained on separate models~\cite{bhattacharya2024joint}. 
Recently, advances in information retrieval show that
unifying these two tasks within a shared model offers significant advantages~\cite{yao2021user,si2023search}: 
First, joint modeling of user behaviors across both S\&R provides a more comprehensive understanding of user preferences. 
Second, leveraging interactions from both task domains enriches item-side interactions, enhancing the understanding and delivery of items. 

\begin{figure}[t]  
\setlength{\abovecaptionskip}{0cm}
\setlength{\belowcaptionskip}{-0.4cm}
    \centering    
    \includegraphics[width=\linewidth]{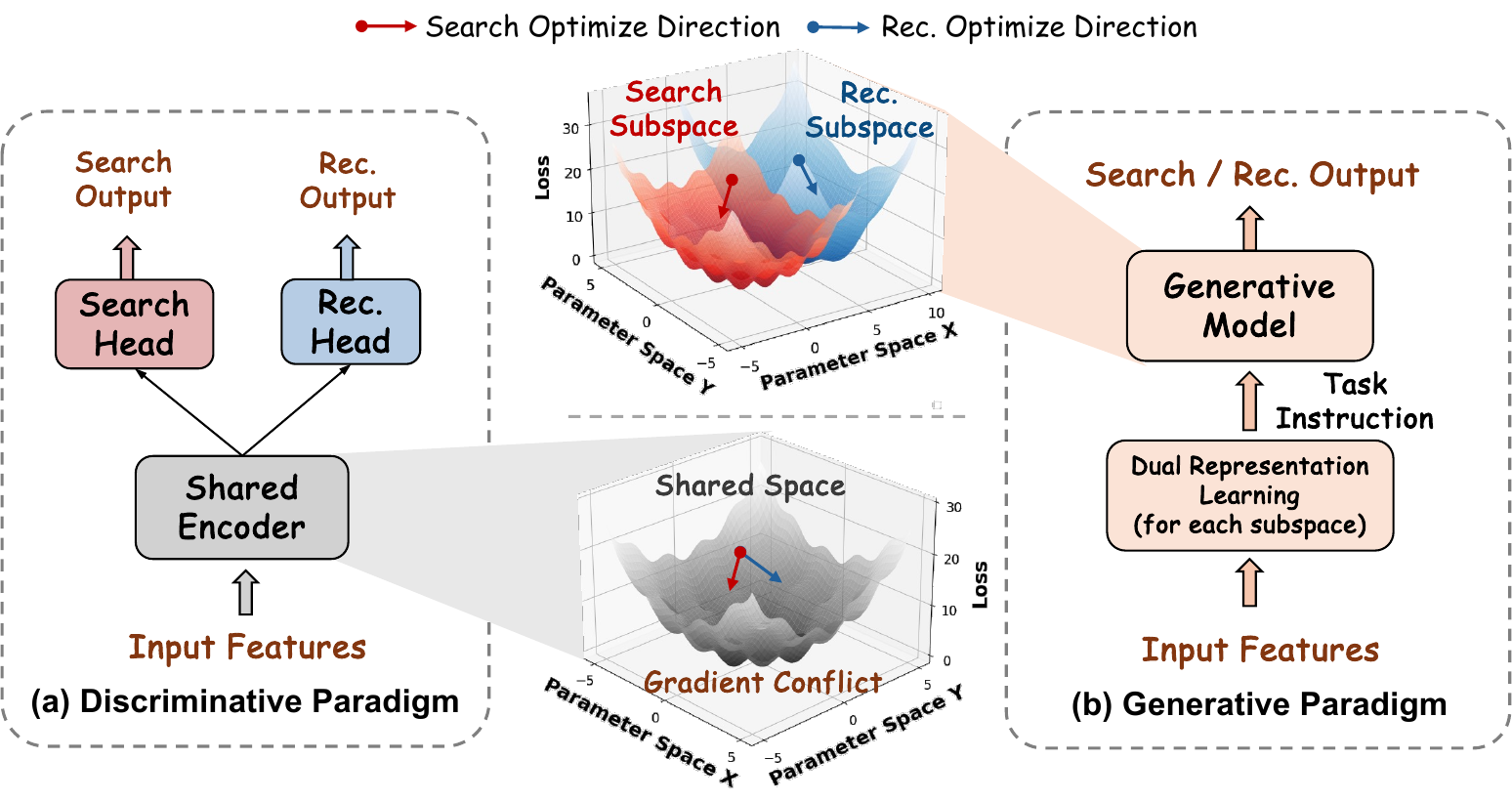}
    \caption{(a) and (b) illustrate the discriminative and generative paradigms for unifying S\&R, respectively, highlighting their differences in parameter space. While the discriminative paradigm results in low mutual information by optimizing S\&R tasks within a shared parameter space, the generative paradigm enhances mutual information by partitioning the space into distinct subspaces through instructions.}
    \label{fig:intro}
\end{figure}

Existing efforts to unify S\&R mostly follow a discriminative paradigm, which utilizes shared encoders to process input features (\eg user interaction history) and task-specific heads to predict whether a user will click on an item in S\&R tasks, as illustrated in Figure~\ref{fig:intro}(a)~\cite{shi2024unisar,xie2024unifiedssr}. 
However, such paradigm has two drawbacks:
1) \textbf{Gradient conflict:}
The differing optimization objectives of S\&R, where search primarily focuses on modeling query-item relevance based on the current query context, while recommendation emphasizes capturing user preferences by analyzing user’s historical interactions, lead to gradient conflicts in the shared encoder~\cite{shi2023recon,he2024multibalance}.
This often results in negative cosine similarity between the task gradients and interferes with effective model optimization~\cite{yu2020gradient} (see empirical evidence in Section~\ref{sec:gradient}).
2) \textbf{Manual architecture design overhead:} 
The design of task-specific heads requires substantial manual effort, increasing system complexity and reducing scalability and extensibility~\cite{rahimian2023dynashare,xie2024unifiedssr}.


Inspired by information theory, we argue that these drawbacks potentially both stem from the same underlying issue --- low mutual information between the model's input features and task-specific output during optimization.
This low mutual information arises from constraining both tasks within a shared parameter space in the shared encoder, as depicted in the parameter space analysis in Figure~\ref{fig:intro}(a).
This constraint limits the representation capability of the shared encoder, resulting in gradient conflict.
Consequently, it necessitates manual designs of task-specific heads to compensate the potential performance degradation caused by gradient conflicts.

To address the above limitations, we aim to increase the mutual information between the model’s input features and the task-specific outputs. 
Partitioning the parameter space into distinct subspaces has proven to be effective for enhancing mutual information, particularly when addressing diverse tasks~\cite{ueda2020latent}.
It reduces task interference and enables the model to learn task-specific predictions through different task optimization subspaces.
To naturally achieve such partitioning, we design a generative paradigm that leverages instruction tuning in pre-trained generative models.
Instruction tuning partitions the parameter space by introducing distinct prompts for each task, effectively tailoring a unique optimization subspace for S\&R as illustrated in Figure~\ref{fig:intro}(b).

To this end, we propose a novel \uline{\textbf{Gen}}erative paradigm for unifying \uline{\textbf{S}}earch and \uline{\textbf{R}}ecommendation, abbreviated as GenSR, which leverages task-specific prompts to partition the model’s parameter space into subspaces, thereby enhancing mutual information (see detailed analysis in \S\ref{sec:theoretical}).
To construct effective subspaces for each task, GenSR first prepares informative representations for each subspace and then optimizes both subspaces in one unified model.
Specifically, GenSR consists of two main modules: 
\begin{enumerate*}[label=(\arabic*)]
\item \emph{Dual representation learning}:
GenSR independently models historical dual representations that encompass sufficient information for both S\&R tasks.
Specifically, we simultaneously learn item representations from both \ac{CF} view and semantic view.
Meanwhile, to capture user preference from extensive S\&R historical interactions, a soft-filtering strategy is designed to prioritize the most relevant representations.
\item \emph{S\&R task unifying}: 
GenSR integrates previously learned representation with task-specific prompts and feeds them into a shared generative model to perform both tasks.
Specifically, it employs contrastive learning to align the CF and semantic representations for the same users and uses instruction tuning to effectively generate task-specific outputs.
\end{enumerate*}
\smallskip
\noindent In summary, our main contributions are as follows:
\begin{enumerate}[label=(\arabic*), leftmargin=*]
\item We propose GenSR, a generative paradigm designed to unify S\&R tasks through dual representation learning and S\&R task unifying.
\item We introduce a novel perspective for comparing discriminative and generative paradigms based on mutual information.
\item We conduct comprehensive experiments on two public datasets, including S\&R performance analysis, mutual information estimation, parameter space visualization, gradient conflict analysis and ablation study of different modules, to validate GenSR’s effectiveness and efficiency compared to other baselines across both S\&R tasks.
\end{enumerate}

\section{Related Work}
\label{sec:related_work}

\subsection{Unifying search and recommendation}
Recent research highlights that search and recommendation (S\&R) behaviors are inherently complementary, as each provides unique signals that enhance user modeling and item understanding~\cite{penha2024bridging,shi2025unified}.
By jointly leveraging data from both tasks, it is possible to achieve more accurate intent modeling and representation learning, thereby improving the performance of both tasks.
As a result, an increasing number of studies have focused on unifying S\&R.
These efforts can be broadly categorized into two main directions:


\textbf{(1) Search-enhanced recommendation}~\cite{wu2019neural}, 
which integrates search behavior into the recommendation scenario to improve user preference modeling. 
This line of research assumes that users’ search queries convey semantic intent signals that are often missing from the behavioral data used in recommendation systems.
For instance, SESRec~\cite{si2023search} utilizes contrastive learning to extract similar and dissimilar user interests from S\&R, and then leverages users' search interests for recommendation.
Query-SeqRec~\cite{he2022query} effectively incorporates explicit user queries from the search side, alongside item interactions, for improved user intent modeling and recommendation performance.


\textbf{(2) Unifying S\&R}~\cite{yao2021user}, which shares S\&R data and utilizes one model to boost the performance on both tasks.
This line of work typically utilizes shared representations or architectures to model S\&R behaviors within a unified framework. 
For example, 
UnifiedSSR~\cite{xie2024unifiedssr} jointly models user behavior history in S\&R scenarios using a parameter-sharing dual-branch network and an intent-oriented session module.
UniSAR~\cite{shi2024unisar} models fine-grained user behavior transitions between S\&R through extraction, alignment, and fusion.

However, they follow the discriminative paradigm, which introduces gradient conflict and manual design complexity. 
Recently, \citet{penha2024bridging} proposed a generative model for unifying S\&R. 
While promising, it treats S\&R data independently, ignoring their joint interaction history—a rich source of contextual signals for user modeling. 
In contrast, we introduce a generative paradigm that jointly leverages S\&R interactions to better model user and item representations and enhances mutual information.


\subsection{Generative recommendation}
Generative models, notably Generative Adversarial Networks (GANs)~\cite{wang2017irgan,gao2021recommender,jin2020sampling} and Variational Autoencoders (VAEs)~\cite{ma2019learning,liang2018variational,zhang2017autosvd++}, are pivotal for personalized recommendations but face structural limitations~\cite{sohl2015deep, kingma2016improved}. 
Nowadays, pre-trained generative models have garnered attention in recommendation for their contextual understanding and global knowledge~\cite{lin2024data}, which better capture user preferences from interactions.
Early studies focused on fine-tuning these models with tailored prompts and recommendation data, e.g., P5~\cite{geng2022recommendation}, TALLRec~\cite{bao2023tallrec}, and M6-Rec~\cite{cui2022m6}. 
For instance, TALLRec introduces an effective and efficient instruction tuning framework to align large language models (LLMs) with recommendation tasks, addressing the challenge that LLMs are not originally pre-trained for such purposes.
Subsequent advancements, notably BIGRec~\cite{bao2023bi}, IDGenRec~\cite{tan2024idgenrec} and TIGER~\cite{rajput2023recommender}, extend this approach by grounding outputs in specific item spaces or incorporating semantic details to refine recommendations. 
For example, BIGRec first fine-tunes LLMs to generate meaningful item descriptions and then maps these to actual items, enhancing recommendation performance.
IDGenRec proposes learning unique, semantically rich textual IDs for items to seamlessly integrate them
into LLM-based generative recommender systems.
Recently, researchers have recognized the critical role of CF information in generative recommendation, as CF plays a crucial role in understanding user and item representations in recommendation systems~\cite{wang2024learnable,kim2024large}.  

However, integrating search data into generative recommendations remains underexplored and holds potential for capturing users’ immediate intent to boost personalization.

\subsection{Personalized search}
Personalized search aims to deliver items tailored to users’ preferences, behaviors, and contextual information.
Early methods learn user representations from profiles or search histories to re-rank items based on personalized relevance.
For instance, HEM learns user representations from user profiles to provide personalized search rankings~\cite{ai2017learning}.
TEM enhances this by incorporating user search histories, allowing it to better capture users’ preferences and historical behaviors~\cite{bi2020transformer}.
Beyond these, some approaches tackle key challenges such as sparse user behavior and noisy interaction data.
For instance, CoPPS~\cite{dai2023contrastive} leverages contrastive learning to enhance user representations under sparse and noisy search scenarios. 
QIN~\cite{guo2023query} incorporates cascaded filtering to extract query-relevant behaviors and employs fused attention to distill salient signals from long-tail history.
However, most personalized search methods rely solely on search-side signals, overlooking the rich collaborative filtering information available in recommendation interactions.
This limits their ability to effectively model long-term user preferences.

In contrast, our proposed model adopts a unified framework that integrates both recommendation and search behaviors within a generative paradigm, enabling the construction of task-aware prompts and richer user representations to improve retrieval quality.

\subsection{Generative model for ranking}
Due to the advanced contextual understanding capabilities of pre-trained language models, generative models for ranking have emerged as a novel approach to directly model the relevance between queries and items by casting ranking as a sequence generation problem. 
These methods approach ranking by generating sequences conditioned on input queries and documents, shifting away from traditional scoring-based frameworks. 
By treating ranking as a text generation task, they are capable of modeling complex interactions between queries and items in a more flexible and contextualized manner.
For example, ~\citet{nogueira2020document} adapts T5 for document ranking by training it to generate relevance labels (``true'' or ``false'') for query-document pairs. 
Similarly, ~\citet{ferraretto2023exaranker} introduces a generative ranking model trained to output both relevance labels and explanations for query-document pairs.

Despite these advances, it's promising to incorporate recommendation data to extract users' preferences which helps to boost the performance on the search side.

\section{Task Formulation}
\label{sec:task_formulation}
We formulate the task of unifying S\&R within both discriminative and generative paradigms.
The goal is to integrate users’ historical S\&R interactions into a single model capable of performing both tasks according to user needs.
In the search setting, the objective is to retrieve documents relevant to a user’s query while leveraging their historical S\&R interactions. 
In the recommendation setting, the objective is to suggest items based on the user’s historical S\&R interactions.
Formally, let $\mathcal{U}$ and $\mathcal{I}$ be the sets of users and items. 
For a user $u \in \mathcal{U}$, the model predicts preference for a candidate item $i_c$ in either a search or recommendation context.
To achieve this, the model takes as input three types of information for each user $u$:
\begin{enumerate}
    \item \textbf{Interaction history} $H_u = \{(i_1, b_1),(i_2, b_2), \cdots,(i_N, b_N)\}$, 
where $i_n\in\mathcal{I}$ is the $n-$th item in the interaction history, 
$N$ is the total number of interactions,
$b_n$ is the type of behavior for $n-$th item, with "src" indicating search and "rec" indicating recommendation.
    \item \textbf{Candidate item} $i_{c}$, which is the item candidate pool used for evaluating whether the user $u$ will like (i.e., click) this item.
    \item \textbf{Query} $q$, issued if the current task is search; otherwise, $q$ is empty for recommendation.
\end{enumerate}

\noindent\textbf{Discriminative paradigm.}
This paradigm encodes inputs via shared encoders and applies task-specific heads. 
The model minimizes the combined cross-entropy loss for both tasks:
\begin{equation}
\label{eqn:task_traditional_loss}
\mathop{\min}_{\Phi} \{\mathcal{L}_{\Phi} = \mathcal{L}{s}(H_u, i_c, q) + \gamma \mathcal{L}{r}(H_u, i_c) \},
\end{equation}
where $\Phi$ denotes model parameters,  $\gamma$ controls the trade-off between S\&R tasks, and
$\mathcal{L}_s$, $\mathcal{L}_r$ are cross-entropy losses for S\&R.

\noindent\textbf{Generative paradigm.}
Differently, in the generative paradigm,  we cast both S\&R as conditional text generation.
The input information is transformed into a unified structured prompt $x$, which is tokenized before being fed into the model for instruction tuning.
The model generates ``Yes'' or ``No'' to indicate user preference via constrained generation.

The training objective minimizes the negative log-likelihood of the users' preference $y$ conditioned on input $x$ in an auto-regressive manner:
\begin{equation}
\label{eqn:task_llm_loss}
    \mathop{\min}_{\Theta} \{\mathcal{L}_{\Theta}=-\sum_{t=1}^{|y|} \log P_{\Theta}(y_t|y_{<t},x)\}, 
\end{equation}
where $\Theta$ denotes model parameters, $x$ and $y$ denote the “Instruction Input” and “Instruction Output” in the self-instruct data for instruction tuning, $y_t$ is the $t$-th token of $y$, and $y_{<t}$ represents all tokens preceding $y_t$.

\section{Mutual Information Analysis of GenSR}
\label{sec:theoretical}
In this section, we conduct an in-depth analysis of mutual information to support our claim that the proposed generative paradigm effectively enhances mutual information compared to traditional discriminative methods. 
We begin by defining mutual information from an information-theoretic perspective, and then present the theoretical evidence to substantiate this claim.

\subsection{Information-theoretic perspective}
\label{sec:info}
From an information-theoretic standpoint, the representational capability of a unified S\&R model can be examined via the mutual information between the input \(X\) and the output \(Y_t\) for a task \(t \in \{S, R\}\). 
Mutual information, denoted as $I(X; Y_t)$, measures how much information the input $X$ provides about the output $Y_t$, quantifying their dependence. 
By definition:
\begin{equation}
  I(X; Y_t) \;=\; H(Y_t) \;-\; H(Y_t \mid X),
  \label{eq:mutual_info_def}
\end{equation}
where \(H(Y_t)\) is the entropy of \(Y_t\), and \(H(Y_t \mid X)\) is the conditional entropy of \(Y_t\) given \(X\). 

A higher mutual information indicates that the input \(X\) contains more predictive information about the output \(Y_t\), reflecting a stronger alignment between inputs and task-specific outputs, i.e., higher model representation capability. 
Conversely, lower mutual information implies that predicted output fail to fully capture the distribution and pattern of input, reflecting a lower model representation capability.
Therefore, mutual information \(I(X; Y_t)\) serves as a good indicator for model representation capability~\cite{bachman2019learning}.

\subsection{Theoretical analysis}
\label{sec:theoretical_analysis}
In this section, we aim to show the effectiveness of our model GenSR, based on the aforementioned mutual information. Specifically, we conduct a theoretical analysis to show this utilizing the following theorem:


\begin{theorem}
\label{the:mi_separability}
Suppose the input distribution $X \sim \mathcal{N}(\mu, \Sigma)$, and the noise $\epsilon \sim \mathcal{N}(0, \sigma_\epsilon^2 I)$ is independent of $X$. 
Consider two paradigms: generative (with parameters $\phi$) and discriminative (with parameter $\theta$). 
Assume the loss functions $\mathcal{L}_\theta$ and $\mathcal{L}_\phi$ admit second-order Taylor expansions around their respective true parameters with bounded remainder terms,\footnote{This assumption is common in statistical learning and information retrieval~\cite{prasad2017separability,kishida2003pseudo}.}
then the mutual information between the input and output satisfies:
\begin{equation}
\label{eqn:theorem3}
I_\theta(X; Y_S) + I_\theta(X; Y_R)
\leq
I_{\phi}(X; Y_S) + I_{\phi}(X; Y_R).
\end{equation}
Here, $I_{\phi}(X; Y_S)$ and $I_{\phi}(X; Y_R)$ denote the mutual information between input and output under the generative paradigm for S\&R tasks, respectively, while $I_\theta(X; Y_S)$ and $I_\theta(X; Y_R)$ denote mutual information under the discriminative paradigm.
\end{theorem}


We provide a detailed proof of Theorem~\ref{the:mi_separability} from the view of parameter approximation separability in the Appendix.
Intuitively, we prove that \textit{generative paradigm can enhance mutual information in S\&R scenario compared with discriminative paradigm under certain input distribution assumption.}
    




\section{Method}
\label{sec:method}

\begin{figure*}
\setlength{\abovecaptionskip}{0.03cm}
\setlength{\belowcaptionskip}{-0.2cm}
\centering
\includegraphics[width=\textwidth]{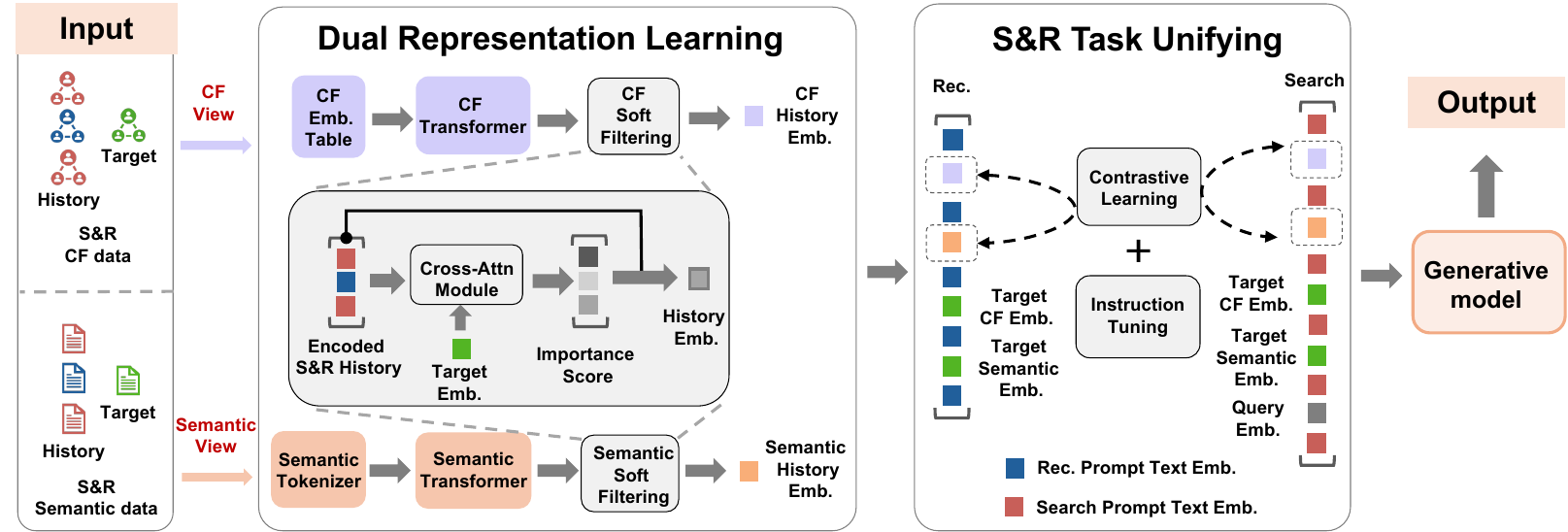}
\caption{GenSR structure. Given S\&R history, GenSR first utilizes dual representation learning to obtain CF and semantic history embeddings. The S\&R task unifying module then integrates these dual representations with task-specific prompts into a shared generative model, leveraging contrastive learning for view alignment and instruction tuning to effectively generate task-specific outputs for both search and recommendation.}
\label{fig:method}
\end{figure*}

In this section, we introduce GenSR, a novel generative paradigm for unifying search and recommendation (S\&R).
GenSR has two modules:
1) Dual Representation learning, which independently learns item representations from both the CF view and the semantic view.
2) S\&R Task Unifying, which utilizes contrastive learning together with instruction tuning to generate task-specific outputs effectively. 
The overall architecture of GenSR is shown in Figure~\ref{fig:method}.

\subsection{\textbf{Dual representation learning}}
\label{sec:dual-view}
To derive informative dual representation for each subspace, GenSR first models historical dual-view representation to encompass sufficient information for both S\&R tasks, and then employs soft-filtering strategy to capture use preference from extensive S\&R historical representations.

\subsubsection{\textbf{Dual-view representation}}
When unifying S\&R tasks, it is crucial to model item representations that encapsulate sufficient information to effectively support both tasks. 
A key consideration in this process is the distinct focus of S\&R tasks on item representations.
Specifically, for the \textbf{search} task, item representations rely heavily on semantic matching between queries and items, as users actively search for specific items based on their explicit information needs. 
In contrast, for the \textbf{recommendation} task, item representations are mainly shaped by CF information derived from historical user interactions~\cite{zhao2023popularity}, as recommendations are passively received by users who often lack explicit information needs. 

This distinction underscores the need to integrate both semantic and CF information to construct comprehensive item representations.
To tackle this challenge, we propose a dual-view representation method that separately captures semantic and CF information from two views.

\noindent\textbf{CF view.}
To model the CF view for the dual-view representation, we begin by extracting the S\&R history $H_u=\{i_1, i_2, \cdots, i_N\}$ for a given user $u$.
This history represents all chronologically ordered S\&R interactions, where $N$ denotes the total number of interactions recorded in the user’s history $H_u$.

From this history, we compute the corresponding CF representations $E_u^\text{cf} =\{e_1, e_2, \cdots, e_N\}^T \in \mathbb{R}^{N \times d}$ using a pre-trained CF embedding table $\mathcal{E} \in \mathbb{R}^{|\mathcal{I}| \times d}$.
Here, each $e_i$ is a $d$-dimensional embedding, $|\mathcal{I}|$ denotes the total number of items in the corpus, and $d$ represents the dimensionality of the CF embeddings.\footnote{The CF embedding table $\mathcal{E}$ is derived from training LightGCN using only the recommendation behaviors in the dataset~\cite{he2020lightgcn}.}

To capture the sequential dependencies within the S\&R history, the representations $E_u^\text{cf}$ are further processed using a transformer encoder. 
The encoder comprises a Multi-Head Self-Attention (MSA) layer and a Feed-Forward Network (FFN). 
Specifically, the MSA layer takes $E_u^\text{cf}$ as the input, where the query ($Q$), key ($K$), and value ($V$) matrices are all set to $E_u^\text{cf}$. Then the CF representation $C_u \in \mathbb{R}^{N \times d}$ is computed as:
\begin{equation}
\label{eqn:cf_transformer}
    C_u = FFN(MSA(E_u^\text{cf},E_u^\text{cf},E_u^\text{cf})), 
\end{equation}

\noindent\textbf{Semantic view.}
To model the semantic view for the dual-view representation, we process the user’s S\&R history  $H_u=\{i_1, i_2, \cdots, i_N\}$, which contains all chronologically ordered interactions, where $N$ is the total number of interactions. 
Semantic representations $E_u^\text{sem} =\{s_1, s_2, \cdots, s_N\}^T \in \mathbb{R}^{N \times d'}$ are derived by tokenizing the textual descriptions of items using a generative model tokenizer. Here, $d'$ is the embedding dimensionality of the generative model, and each $s_i$ is a $d'$-dimensional embedding capturing the item’s semantic information.

To capture the sequential dependencies, the representations $E_u^\text{sem}$ are fed into a transformer encoder.
The transformer encoder consists of a Multi-head Self-Attention (MSA) layer and a Feed-Forward layer (FFN). 
Specifically, MSA takes $E_u^\text{sem}$ as the input, where $Q=K=V=E_u^\text{sem}$. 
Then the semantic representation $S_u \in \mathbb{R}^{N \times d'}$ is computed as: 
\begin{equation}
\label{eqn:sem_transformer}
    S_u = FFN(MSA(E_u^\text{sem},E_u^\text{sem},E_u^\text{sem})).
\end{equation}

\subsubsection{\textbf{S\&R history soft filtering}}
\label{sec:soft-filter}
When combining S\&R behaviors together in the user history sequence, it is essential to effectively capture user preferences from an extensive history of S\&R interactions. 
The fusion of these two tasks introduces a larger volume and greater complexity of user behaviors, making it crucial to extract meaningful patterns from the multi-task interaction history. 
To address this challenge, we propose a \textit{history soft filtering strategy}, which selectively soft-filters relevant items from the extensive S\&R interaction history based on the user’s specific information needs.


Intuitively, after obtaining the learned CF representation $C_u$ and semantic representation $S_u$ from the dual-view representation, we calculate the importance score of each historically interacted item with respect to the candidate item $i_c$. 
Specifically, for the CF side, given a user $u$, the CF representation $C_u = \{h_1, h_2, \dots, h_N\}$ contains the encoded CF embeddings of $N$ historically interacted items, where $h_j$ is the encoded CF embedding of the $j$-th item.
To determine the relevance of each $h_j$ to the candidate item $i_c$ with CF embedding $e_{i_c}$, we compute the importance score $\alpha_j$ using a multi-head cross-attention (MCA) module:
\begin{equation}
\label{eqn:importance-score}
\alpha_j = \exp\left(\text{MCA}(e_{i_c}, h_j)\right) / \sum_{k=1}^N \exp\left(\text{MCA}(e_{i_c}, h_k)\right),
\end{equation}
where $\alpha_j$ is the importance score for the $j$-th historical item, 
and $\text{MCA}(e_{i_c}, h_j)$ represents the attention score between the candidate item’s CF embedding  $e_{i_c}$ and the $j$-th item CF embedding $h_j$.  
Through the MCA module, the candidate item \(e_{i_c}\) dynamically queries and weights the most relevant historical item embeddings \(h_j\), providing importance signals for subsequent soft filtering.

We then apply soft-filtering to the S\&R interaction history using the calculated importance scores $\alpha_j$:
\begin{equation}
\label{eqn:soft-filter}
C_u^{\text{filter}} = \sum_{j=1}^N \alpha_j \cdot W_V h_j,
\end{equation}
where $W_V$ is a learned projection matrix 
associated with $C_u$, and $h_j$ is the $j$-th optimized CF embedding from $C_u$.

Similarly, we perform the same soft-filtering for the semantic side, obtaining $S_u^{\text{filter}}$. By learning both the filtered CF and semantic representations of historical S\&R interactions, we obtain relevant dual representation that supports both S\&R tasks.

\subsection{\textbf{S\&R task unifying}}
\label{sec:framework}
After preparing the dual representation, GenSR integrates these learned representation with task-specific prompts, which are then fed into a shared generative model to unify S\&R tasks.

\subsubsection{\textbf{Training}}
\label{sec:training}
We use instruction tuning to train the generative model for performing S\&R tasks, enabling the parameter space to naturally partition into subspaces optimized for different tasks.
Specifically, after obtaining $C_u$ and $S_u$ from the dual representation learning, we construct a task-specfic prompt that incorporates these representations, along with the query $q$ (for search tasks only), the candidate item’s CF embeddings $e_{i_c}$ and semantic embeddings $s_{i_c}$.
Formally, we use generative model tokenizer to tokenize the textual content of the prompt, which contains textual instructions and the query (if applicable). 
The result of the tokenization is represented as:
$P_t = [p_{1}^t, \langle PH1 \rangle, p_{2}^t, \langle PH2 \rangle, p_{3}^t, \langle PH3 \rangle, p_{4}^t, \langle PH4 \rangle, p_{5}^t]$,
where task $t \in \{S,R\}$, $p_{k}^t$ denotes $k$-th segments of prompt text tokens for task $t$, and $\langle PH \rangle$ represents placeholders that will be replaced with other feature embeddings later.\footnote{The five segments of prompt text tokens ($p_{1}^t$ to $p_{5}^t$) are used because, in our setting, we only have four feature embeddings to insert into the prompt. Therefore, the prompt is divided into five segments to interleave the embeddings properly.}
We then encode the prompt into a sequence of embeddings and replace the placeholders with learned representation and candidate item embeddings for each task:
\begin{equation}
\label{eqn:emb}
E_t = [e_{{1}}^t, C_u, e_{{2}}^t, S_u, e_{{3}}^t, e_{i_c}, e_{{4}}^t, s_{i_c}, e_{{5}}^t], \quad \text{where} \quad t \in \{S, R\}.
\end{equation}
Here, $e_{k}^t \in \mathbb{R}^{1 \times d_2}$ represents the embedding for token $p_1^k$, obtained via the embedding lookup in the generative model.
The final tokenized embedding $E_t$ is fed into the generative model as input for instruction tuning. 
This approach allows the model to adaptively leverage both CF and semantic information, while employing task-specific prompts to partition the model’s parameter space into subspaces for optimizing different tasks.

\subsubsection{\textbf{Contrastive learning between CF and Semantic views}}
\label{sec:contrastive-learning}
Since we design dual representation learning from the CF and semantic views, it is important to notice that these perspectives provide complementary insights into user preferences.
Aligning these two views enables the model to integrate their respective strengths, resulting in comprehensive and robust user representations that enhance generalization across diverse S\&R tasks.

To effectively align user representations from both the CF and semantic views, we employ contrastive learning.
This approach encourages the CF and semantic representations of the same user $u$ to be closer in the embedding space, while ensuring that the CF or semantic representations of different users $u'$ (within the same batch) are pushed further apart. 

Specifically, for a given user $u$, with CF representation $C_u$ and semantic representation $S_u$ obtained from dual-view representation, the goal is twofold:
1) Minimize the distance between $C_u$ and $S_u$ (positive pair), ensuring that the CF and semantic embeddings of the same user are closely aligned, effectively transferring complementary information between views.
2) Maximize the distance between $C_u$ (or $S_u$) and the representations $C_{u'}$  or $S_{u{\prime}}$ of other users $u'$  in the batch (negative pairs).
This enforces distinctiveness between users, preventing overlapping embeddings that could reduce model performance in S\&R tasks. 
The contrastive loss for this alignment is calculated as follows:
\begin{equation}
\label{eqn:cl}
\scalebox{0.88}{%
$\mathcal{L}_c = -\frac{1}{|B|} \sum\limits_{u \in B} \log \frac{\exp(\text{sim}(C_u, S_u) / \tau)}{\sum_{u^\prime \in B} \exp(\text{sim}(C_u, S_{u^\prime}) / \tau) + \sum_{u^\prime \in B} \exp(\text{sim}(S_u, C_{u^\prime}) / \tau)}$%
},
\end{equation}
where $C_u$ and $S_u$ denote the CF and semantic embeddings for the same user $u$, while $u'$ refers to other users in the batch $B$. The function $\text{sim}(\cdot, \cdot)$ is implemented as cosine similarity, measuring the similarity between two embeddings. 
$\tau$ is a temperature factor that adjusts the sharpness of the similarity distribution, controlling how much emphasis is placed on high-similarity pairs.


\subsubsection{\textbf{Loss function}}
\label{sec:loss}
The overall loss function for the GenSR model consists of two components: 
1) a task-specific generation loss for S\&R, $\mathcal{L}_{\text{S\&R}}$, and 
2) a contrastive learning loss, $\mathcal{L}_c$, which aligns the CF and semantic representations.
The combined loss is:
\begin{equation}
\label{eqn:loss}
\mathcal{L} = \mathcal{L}_{\text{S\&R}} + \beta \mathcal{L}_c, \text{where}
\end{equation}
\begin{equation}
\label{eq:l4}
\mathcal{L}_{\text{S\&R}} =-\sum_{t=1}^{|y|} \log P_{\Theta}(y_t|y_{<t},E)\}, 
\end{equation}
where $\beta$ is the loss balance factor that controls the weight of the contrastive loss relative to the S\&R task loss, and $|y|$ is the length of the target sequence, $y_t$ is the token at position $t$ in the target sequence, and $y_{<t}$ represents all tokens preceding $t$. 
The conditional probability $P_{\Theta}(y_t \mid y_{<t}, E)$ is computed by the generative model, parameterized by overall model parameters $\Theta$, given the input tokenized embedding $E$. 
Intuitively, generation loss $\mathcal{L}_{\text{S\&R}}$ encourages the model to maximize the likelihood of generating the correct output sequence for S\&R.
\section{Experiments}
\label{sec:experiment}

In this section, we conduct a comprehensive experimental study to address the following research questions.

\begin{enumerate}[leftmargin=*]
    \item [-] \textbf{RQ1:} How does the performance of GenSR compare to other baseline methods across various datasets in both recommendation and search scenarios?
    \item [-] \textbf{RQ2:} How well can GenSR enhance mutual information compared to discriminative approaches?
    \item [-] \textbf{RQ3:} How effectively does GenSR mitigate gradient conflict relative to discriminative approaches?
    \item [-] \textbf{RQ4:} What is the impact of different strategies (\eg dual-view representation and S\&R history soft filtering) within GenSR on its overall performance?
\end{enumerate}

\begin{table}[t!]
\setlength{\abovecaptionskip}{0cm}
\setlength{\belowcaptionskip}{0cm}
\caption{Statistics of three datasets. \#Inter-S and \#Inter-R denote the interaction number of search behavior and recommendation behavior, respectively.}
\label{tab:datasets}
\begin{tabular}{cccccc}
\toprule
\textbf{Dataset}     & \textbf{\#User} & \textbf{\#Item} & \textbf{\#Queries} & \textbf{\#Inter-S} & \textbf{\#Inter-R} \\ \midrule 
\textbf{Amazon}    & 68,223         & 61,934          & 4,298               & 934,664                       & 989,618         \\
\textbf{KuaiSAR} & 25,877         & 6,890,707         & 453,667               & 5,059,169                         & 14,605,716         \\
\bottomrule
\end{tabular}
\end{table}

\begin{table*}[t]
\setlength{\abovecaptionskip}{0.3cm}
\setlength{\belowcaptionskip}{0cm}
\caption{Overall performance of GenSR and other baselines under recommendation scenario. Bold signifies the best performance among the baselines and GenSR while underline represents the second-best method. * denotes statistically significant improvements of GenSR over the best baseline, according to t-tests with a significance level of $p$ < 0.01.}
\label{tab:main_exp_rec}
\centering
\resizebox{\textwidth}{!}{ 
\begin{tabular}{ll|cccc|cccc}
\toprule
\multicolumn{2}{c|}{}                                                                & \multicolumn{4}{c|}{\textbf{Amazon}}                                                                                                                              & \multicolumn{4}{c}{\textbf{KuaiSAR}}                                                                                                                              \\
\multicolumn{2}{c|}{\multirow{-2}{*}{\textbf{Method}}}                               & \textbf{R@5}                      & \textbf{R@10}                     & \textbf{N@5}                        & \textbf{N@10}                       & \textbf{R@5}                      & \textbf{R@10}                     & \textbf{N@5}                        & \textbf{N@10}                       \\ \midrule
\multicolumn{1}{l|}{}                                          & \textbf{DIN~\cite{zhou2018deep}}        & 0.5170                                  & 0.6525                                 & 0.3726                                 & 0.4165                                 & 0.4509                                 & 0.6179                                 & 0.3104                                 & 0.3643                                 \\
\multicolumn{1}{l|}{}                                          & \textbf{GRU4Rec~\cite{hidasi2015session}}    & 0.4949                                 & 0.6548                                 & 0.3388                                 & 0.3907                                 & 0.3764                                 & 0.5788                                 & 0.2345                                 & 0.3087                                 \\
\multicolumn{1}{l|}{}                                          & \textbf{SASRec~\cite{kang2018self}}     & 0.5295                                 & 0.6772                                 & 0.3747                                 & 0.4225                                 & 0.4065                                 & 0.6007                                 & 0.2671                                 & 0.3298                                 \\
\multicolumn{1}{l|}{}                                          & \textbf{BERT4Rec~\cite{sun2019bert4rec}}   & 0.5311                                 & 0.6658                                 & 0.3954                                 & 0.4390                                  & 0.3699                                 & 0.5885                                 & 0.2381                                 & 0.3083                                 \\
\multicolumn{1}{l|}{}                                          & \textbf{FMLP-Rec~\cite{zhou2022filter}}   & 0.5356                                 & 0.6879                                 & 0.3739                                 & 0.4232                                 & 0.4292                                 & 0.6159                                 & 0.2851                                 & 0.3453                                 \\
\multicolumn{1}{l|}{}                                          & \textbf{CoLLM~\cite{zhang2023collm}}   &   0.5800                               &   0.6811                              &                    0.4651              &     0.5001                             &        0.4092                         &        0.5915                      &         0.3199                        &          0.3683                      \\
\multicolumn{1}{l|}{\multirow{-6}{*}{\textbf{Recommendation}}} & \textbf{IDGenRec~\cite{tan2024idgenrec}}   & 0.2305                                 & 0.2818                                 & 0.1791                                 & 0.1955                                 & -                                      & -                                      & -                                      & -                                      \\ \midrule
\multicolumn{1}{l|}{}                                          & \textbf{NRHUB~\cite{wu2019neural}}     & 0.4988                                 & 0.6503                                 & 0.3487                                 & 0.3977                                & 0.3862                                 & 0.5610                                 & 0.2572                                 & 0.3136   \\
\multicolumn{1}{l|}{}                                          & \textbf{Query-SeqRec~\cite{he2022query}}     & 0.5401                                 & 0.6860                                 & 0.3859                                 & 0.4333                                & 0.3920                                 & 0.5890                                 & 0.2552                                 & 0.3186   \\
\multicolumn{1}{l|}{}                                          & \textbf{SESRec~\cite{si2023search}}     & 0.5623                                 & 0.6864                                 & 0.4245                                 & 0.4648                                 & 0.4956                                 & 0.6643                                 & 0.3432                                 & 0.3978                                 \\
\multicolumn{1}{l|}{}                                          & \textbf{JSR~\cite{zamani2018joint}}        & 0.5467                                 & 0.6779                                 & 0.3970                                  & 0.4396                                 & 0.4791                                 & 0.6453                                 & 0.3315                                 & 0.3853                                 \\
\multicolumn{1}{l|}{}                                          & \textbf{USER~\cite{yao2021user}}       & 0.5441                                 & 0.6854                                 & 0.3964                                 & 0.4422                                 & 0.4086                                 & 0.5627                                 & 0.2820                                  & 0.3318                                 \\
\multicolumn{1}{l|}{}                                          & \textbf{UnifiedSSR~\cite{xie2024unifiedssr}} & 0.5196                                 & 0.6707                                 & 0.3662                                 & 0.4151                                 & 0.3981                                 & 0.5939                                 & 0.2617                                 & 0.3249                                 \\
\multicolumn{1}{l|}{}                                          & \textbf{UniSAR~\cite{shi2024unisar}}     & \uline{0.5874}                                 & \uline{0.7020}                                  & \uline{0.4513}                                 & \uline{0.4885}                                 & \uline{0.5169}          & \uline{0.6792}          & \uline{0.3632}          & \uline{0.4158}          \\
\multicolumn{1}{l|}{}                                          & \textbf{BSR~\cite{penha2024bridging}}     &      0.1417                            &      0.2278                             &         0.0899                         &            0.1177                      &   0.2180        &     0.3395      &     0.1472      &      0.1860     \\
\multicolumn{1}{l|}{\multirow{-8}{*}{\textbf{Joint S\&R}}}       & \textbf{GenSR}      & {\textbf{0.6102*}}& {\textbf{0.7200*}} & {\textbf{0.4922*}} & {\textbf{0.5278*}} & {\textbf{0.5235*}} & {\textbf{0.6819*}} & {\textbf{0.3686*}} & {\textbf{0.4199*}} \\ \bottomrule
\end{tabular}
}
\end{table*}
\begin{table*}[t]
\setlength{\abovecaptionskip}{0.3cm}
\setlength{\belowcaptionskip}{0cm}
\caption{Overall performance of GenSR and other baselines under search scenario. Bold signifies the best performance among the baselines and GenSR while underline represents the second-best method. * denotes statistically significant improvements of GenSR over the best baseline, according to t-tests with a significance level of $p$ < 0.01.}
\label{tab:main_exp_src}
\centering
\resizebox{\textwidth}{!}{ 
\begin{tabular}{@{}ll|cccc|cccc@{}}
\toprule
\multicolumn{2}{c|}{}                                                                                                          & \multicolumn{4}{c|}{\textbf{Amazon}}                                                                                                                              & \multicolumn{4}{c}{\textbf{KuaiSAR}}                                                                                                                              \\
\multicolumn{2}{c|}{\multirow{-2}{*}{\textbf{Method}}}                                                                         & \textbf{R@5}                      & \textbf{R@10}                     & \textbf{N@5}                        & \textbf{N@10}                       & \textbf{R@5}                      & \textbf{R@10}                     & \textbf{N@5}                        & \textbf{N@10}                       \\ \midrule
\multicolumn{1}{l|}{}                                    & \textbf{QEM~\cite{ai2019zero}}        & 0.7100                                   & 0.8186                                 & 0.5066                                 & 0.5422                                 & 0.6020                                  & 0.7182                                 & 0.4575                                 & 0.4953                                 \\
\multicolumn{1}{l|}{}                                    & \textbf{HEM~\cite{ai2017learning}}        & 0.6778                                 & 0.8267                                 & 0.4736                                 & 0.5221                                 & 0.6505                                 & 0.7653                                 & 0.5029                                 & 0.5400                                   \\
\multicolumn{1}{l|}{}                                    & \textbf{AEM~\cite{ai2019zero}}        & 0.7095                                 & 0.8443                                 & 0.5114                                 & 0.5554                                 & 0.5956                                 & 0.7182                                 & 0.4415                                 & 0.4812                                 \\
\multicolumn{1}{l|}{}                                    & \textbf{ZAM~\cite{ai2019zero}}        & 0.7109                                 & 0.8468                                 & 0.5147                                 & 0.5590                                  & 0.6117                                 & 0.7344                                 & 0.4560                                  & 0.4959                                 \\
\multicolumn{1}{l|}{}                                    & \textbf{TEM~\cite{bi2020transformer}}        & 0.8185                                 & 0.9051                                 & 0.6303                                 & 0.6587                                 & 0.6502                                 & 0.7632                                 & 0.4887                                 & 0.5254                                 \\
\multicolumn{1}{l|}{\multirow{-6}{*}{\textbf{Search}}}   & \textbf{CoPPS~\cite{dai2023contrastive}}      & 0.8169                                 & 0.9051                                 & 0.6281                                 & 0.6570                                  & 0.6616                                 & 0.7707                                 & 0.4977                                 & 0.5331                                 \\ \midrule
\multicolumn{1}{l|}{}                                    & \textbf{JSR~\cite{zamani2018joint}}        & 0.7038                                 & 0.8225                                 & 0.5173                                 & 0.5563                                 & 0.7162                                 & 0.7961                                 & 0.5962                                 & 0.6221                                 \\
\multicolumn{1}{l|}{}                                    & \textbf{USER~\cite{yao2021user}}       & 0.7631                                 & 0.8697                                 & 0.6000                                    & 0.6348                                 & 0.7304                                 & 0.8149                                 & 0.6069                                 & 0.6342                                 \\
\multicolumn{1}{l|}{}                                    & \textbf{UnifiedSSR~\cite{xie2024unifiedssr}} & 0.7744                                 & 0.8812                                 & 0.5847                                 & 0.6196                                 & 0.7377                                 & 0.8320                                  & 0.5991                                 & 0.6297                                 \\
\multicolumn{1}{l|}{}                                    & \textbf{UniSAR~\cite{shi2024unisar}}     & \uline{0.8190}           & \uline{0.8977}          & \uline{0.6875}          & \uline{0.7132}          & \uline{0.7476}          & \uline{0.8369}          & \uline{0.6417}          & \uline{0.6708}           \\        
\multicolumn{1}{l|}{}                                          & \textbf{BSR~\cite{penha2024bridging}}     &       0.4786                           &          0.6015                         &       0.3485                           &          0.3883                        &     0.1573      &    0.2075       &   0.1359        &     0.1520      \\
\multicolumn{1}{l|}{\multirow{-5}{*}{\textbf{Joint S\&R}}}       & \textbf{GenSR}   & { \textbf{0.9782*}}          & { \textbf{0.9853*}}          & { \textbf{0.9664*}}          & { \textbf{0.9687*}}          & { \textbf{0.7661*}}          & { \textbf{0.8372}}          & { \textbf{0.7564*}}          & { \textbf{0.7655*}}           \\ \bottomrule
\end{tabular}
}
\end{table*}

\subsection{Experimental setups}
\label{sec:setting}

\subsubsection{\textbf{Datasets.}}  \label{sec:datasets}
We evaluate GenSR on the two publicly available datasets that contain both S\&R histories. 
\begin{enumerate}
\item \textbf{Amazon}\footnote{\url{https://jmcauley.ucsd.edu/data/amazon/}.} is a publicly released product review dataset collected from Amazon.
It contains extensive user-generated product reviews and metadata, such as product titles and descriptions.
We use the "Kindle Store" 5-core sub-dataset from the dataset, 
which contains user reviews and ratings for Kindle books.
Adopting the procedure from prior work~\cite{si2023search,shi2024unisar}, we first generate synthetic search behaviors and queries based on this recommendation dataset and then apply a leave-one-out strategy for dataset splitting.
The detailed process for generating search behaviors is described in~\cite{ai2017learning}. 
Specifically, queries are extracted from the hierarchical category information of each item and paired with user purchase behaviors to construct user–query–item triples. 
Due to the limited availability of real-world unified search and recommendation datasets, this automatically generated search dataset has been widely adopted in prior research on unifying search and recommendation~\cite{si2023search,shi2024unisar,ai2017learning,van2016learning,guo2019attentive}.
\item \textbf{KuaiSAR}\footnote{\url{https://kuaisar.github.io/.}} is a micro-video dataset that offers authentic S\&R behaviors of users, featuring detailed item categories and real user queries.
We preprocess the data to include only users showing both S\&R behaviors. 
Following~\cite{shi2024unisar}, we filter for users and items with at least five interactions and allocate the latest day’s data for testing, the previous day for validation, and the earlier data for training.
\end{enumerate}


The statistics of datasets are shown in Table~\ref{tab:datasets}.

\subsubsection{\textbf{Baselines.}} 
We compare our method with the state-of-the-art recommendation, search, and unifying methods.

The recommendation baselines are:
\begin{itemize}
    \item  \textbf{DIN}~\cite{zhou2018deep} employs a local activation unit that adaptively learns user interest representations, weighting historical behaviors based on their relevance to a candidate ad.
    \item  \textbf{GRU4Rec}~\cite{hidasi2015session} utilizes RNNs for session-based recommendations. It models sequential user interactions within short sessions, improving recommendations where user profiles are absent.
    \item  \textbf{SASRec}~\cite{kang2018self} introduces a self-attention based model for sequential recommendation, adeptly capturing long-term user behavior by adaptively focusing on relevant past items.
    \item  \textbf{BERT4Rec}~\cite{sun2019bert4rec} employs deep bidirectional self-attention with a Cloze objective to learn richer item representations by leveraging both left and right contexts in user behavior sequences.
    \item  \textbf{FMLP}~\cite{zhou2022filter} uses an all-MLP model that employs learnable filters in the frequency domain to effectively denoise sequential user behavior data.
    \item  \textbf{CoLLM}~\cite{zhang2023collm} integrates collaborative embeddings derived from external traditional recommender models into LLMs by mapping them to the LLM's input token space.
    \item \textbf{IDGenRec}~\cite{tan2024idgenrec} proposes learning unique, semantically rich textual IDs for items to seamlessly integrate them into LLM-based generative recommender systems.\footnote{We only implement IDGenRec on Amazon due to its requirement for explicit semantic textual information of items, which is unavailable for KuaiSAR.}
\end{itemize}

The search baselines are:
\begin{itemize}
    \item  \textbf{QEM}~\cite{ai2017learning} learns a dense vector representation for a given search query, allowing for semantic similarity calculations with products in the same embedding space.
    \item  \textbf{HEM}~\cite{ai2017learning} jointly learns distributed representations for words, queries, users, and products in a unified latent space, effectively personalizing product search by composing user and query embeddings.
    \item  \textbf{AEM}~\cite{ai2019zero} is an attention-based model that personalizes product search by dynamically weighting items in a user's purchase history according to the current query.
    \item  \textbf{ZAM}~\cite{ai2019zero} introduces a novel Zero Attention Strategy, allowing the model to automatically determine when and how much to personalize by optionally attending to a zero vector, thus enabling differentiated personalization.
    \item  \textbf{TEM}~\cite{bi2020transformer} leverages a transformer architecture to dynamically control the influence of personalization and model the interactions between purchased items for personalized product search.
    \item  \textbf{CoPPS}~\cite{dai2023contrastive} introduces contrastive learning to learn robust user representations for personalized product search by leveraging diverse data augmentation strategies that exploit intrinsic sequence correlations and external knowledge graphs to overcome data sparsity and noise.
\end{itemize}

The unifying baselines include two categories:

(a) Search enhanced recommendation:
\begin{itemize}
    \item  \textbf{NRHUB}~\cite{wu2019neural} enriches user representations by integrating heterogeneous user behaviors (search queries, browsed webpages, and clicked news) through an attentive multi-view learning framework.
    \item  \textbf{Query-SeqRec}~\cite{he2022query} leverages previously ignored user queries as a crucial contextual signal to enhance sequential recommendation accuracy by better understanding evolving user intent
    \item  \textbf{SESRec}~\cite{si2023search} enhances sequential recommendation by learning disentangled representations of similar and dissimilar user interests from their combined search and recommendation behaviors.
\end{itemize}

(b) Jointly unified S\&R:
\begin{itemize}
    \item  \textbf{JSR}~\cite{zamani2018joint} jointly models search and recommendation tasks by learning shared item representations and minimizing a combined loss function.
    \item  \textbf{USER}~\cite{yao2021user} integrates heterogeneous user search and recommendation behaviors into a single chronological sequence to learn comprehensive user interests and leverage the interdependencies between the two tasks.
    \item  \textbf{UnifiedSSR}~\cite{xie2024unifiedssr} leverags cross-scenario and cross-view information sharing and dynamically modeling user intent through self-supervised session discovery.
    \item \textbf{UniSAR}~\cite{shi2024unisar} introduces a novel framework to explicitly extract, align, and fuse various fine-grained user transition behaviors between search and recommendation.
    \item \textbf{BSR}~\cite{penha2024bridging} employs atomic item IDs to jointly train generative models for search and recommendation.
\end{itemize}

\subsubsection{\textbf{Experimental settings.}} 
Following previous work~\cite{shi2024unisar,xie2024unifiedssr,yao2021user}, we randomly sample 99 negative items that the user has not interacted with and combine them with the ground truth item to create the candidate item list.  
All the models are then tasked with ranking these candidate items to evaluate top-K S\&R performance using Recall@K and NDCG@K with $K=\{5, 10\}$.
For all the baselines, the hyper-parameters are searched according to the reporting range in the original papers.
To ensure a fair comparison, we implement constrained beam search during inference for two full-ranking generative baselines, IDGenRec and GenSR, which guarantees that the generated results are restricted to the valid candidate set.
Regarding our proposed GenSR, we select flan-t5-base as the backbone.
The best hyper-parameters are selected with the searching scopes as follows:
the temperature factor $\tau$ and the loss balance factor $\beta$ are tuned with the ranges $\{0.01, 0.03, 0.05, 0.07, 0.09\}$ and $\{0.001, 0.01, 0.1, 0.15, 0.2, 0.25, 0.3\}$. 

\subsection{Overall performance (RQ1)}
\label{sec:overall_performance}
We conducted a comprehensive experimental comparison of GenSR’s performance against various baseline methods in both recommendation and search scenarios. As shown in Table~\ref{tab:main_exp_rec} and Table~\ref{tab:main_exp_src}, the experimental results reveal several key findings:

\begin{enumerate}[leftmargin=*]
\item \textbf{Superiority of GenSR}: 
GenSR consistently outperforms all baseline methods across all datasets in both S\&R tasks, achieving state-of-the-art results.
This good performance is primarily attributed to GenSR's generative paradigm and its two core modules: 
Dual Representation Learning and S\&R Task Unifying. 
As discussed in Section~\ref{sec:theoretical}, the generative paradigm enhances the mutual information between input features and task outputs by constructing effective task-specific subspaces. 
This approach mitigates the gradient conflict issues commonly observed in traditional discriminative multi-task learning, as further illustrated in Sections~\ref{sec:visual} and Sections~\ref{sec:gradient}. 
After that, by preparing effective dual-view representations for each subspace and jointly optimizing them within one unified model,
GenSR improves the performance of both tasks while fully leveraging the capabilities of a shared pre-trained model.

\item \textbf{General Advantages of the Unified S\&R Approach}: 
We observe that unified S\&R methods consistently outperform models designed for a single task, both recommendation-only models (e.g., DIN and SASRec) and search-only models (e.g., QEM and TEM). 
These findings provide strong evidence for the intrinsic value of integrating S\&R within a shared framework. 
User behaviors in search and recommendation are not isolated but complementary, with each offering distinct and informative signals. 
Joint modeling enables a more comprehensive understanding of user preferences and facilitates richer item-side interaction signals, ultimately leading to mutual performance gains across both tasks.

\item \textbf{Analysis of GenSR’s SOTA Performance on the Amazon Search Task}: 
Notably, GenSR achieves particularly strong results on the Amazon dataset’s search task. 
This can be explained by the dataset’s nature: 
search queries in Amazon are synthesized from item descriptions, indicating a high degree of alignment and strong semantic correlation between queries and item content. Note that the use of synthetic queries for this dataset to use it for search purposes is in line with prior work (see Section ~\ref{sec:datasets}).
Thanks to GenSR's dual representation learning module, particularly the semantic-view, GenSR is exceptionally capable of capturing and utilizing such fine-grained semantic matching. 
When queries are semantically aligned with item descriptions, GenSR’s generative capabilities allow it to more precisely infer user intent and retrieve relevant items, thereby yielding remarkable performance improvements. 
This highlights the generative paradigm’s strong potential in leveraging semantic information, especially on semantically rich and well-aligned datasets.

\item \textbf{Limitations of Existing Baseline Methods}: 
Not all generative or unified models can achieve GenSR’s performance level. 
For example, IDGenRec performs relatively poorly, especially when compared with GenSR. 
This is likely due to its heavy reliance on semantic textual information to generate unique item IDs, which may hinder its ability to fully capture and utilize traditional CF signals in recommendation tasks. 
In many cases, pure semantic matching is insufficient to reflect complex user-item collaborative preferences. 
GenSR addresses this issue through dual representation learning.
Similarly, BSR exhibits significantly limited performance. 
By generating atomic item IDs, BSR lacks enriched semantic information to enhance item representations. 
Relying solely on atomic IDs may result in poor modeling of item properties and deep user preferences, especially when semantic matching is crucial for search tasks.

\end{enumerate}

\subsection{Potential Application to Full Ranking (RQ1)}
\label{sec:full-rank}
Full ranking represents a highly challenging yet critical evaluation setting for real-world S\&R systems. 
Unlike the more common candidate re-ranking evaluation scenario in unified S\&R, where models operate on a small, pre-selected set of candidate items, full-ranking requires identifying the most relevant items directly from the entire corpus of available items. 
Success in full-ranking demonstrates a model’s robust capability for open-vocabulary item discovery and global relevance estimation, pushing the limits of what search engines and recommender systems can achieve.

A key advantage of the generative paradigm in S\&R lies in its inherent ability to perform full-ranking, seamlessly unifying the retrieval and re-ranking stages into a single, end-to-end optimization framework~\cite{sun2023learning}. 
Unlike traditional two-stage pipelines which first retrieve a subset of candidates and subsequently re-rank them, a generative model directly produces the final ranked items in a single pass. 
This unified approach not only simplifies the overall architecture but also allows the model to jointly optimize retrieval and re-ranking without relying on intermediate heuristic filtering. 
To comprehensively assess the global ranking capability of this generative paradigm, we further evaluate GenSR in the full-ranking setting.

To adapt GenSR for the full-ranking task, we slightly modify its input and output strategy which is suitable for the more typical evaluation setting in unified S\&R.
In the standard evaluation setup, GenSR is typically given a candidate item \(i_c\) along with user history and query, and it generates token probability to indicate preference. 
However, in the full-ranking scenario, the model's goal shifts from assessing a pre-existing candidate to identifying the most relevant item from the entire corpus.
To accommodate this, the input prompt to the generative model is adjusted by excluding the embeddings of the candidate item  (\(e_{i_c}, s_{i_c}\)). 
Instead, the model is provided only with the user's interaction history \(H_u\) and query \(q\) (for search tasks), and it is trained to directly generate the relevant item ID through dual representation learning (excluding soft-filtering strategy due to the lack of candidate item) and S\&R task unifying.

This adjustment transforms the task into an item ID generation problem, similar to current generative retrieval and recommendation methods. 
During inference, we employ constrained beam search, which ensures that the output consists of valid item IDs from the entire corpus. This approach effectively performs full-ranking, enabling the model to discover the most relevant items without relying on a pre-selected candidate set.

Crucially, this adaptation strategy aligns the computational complexity of GenSR for full-ranking with that of existing typical generative retrieval models~\cite{wang2021learning,penha2024bridging}. 
The primary computational burden lies in the decoding phase, where the model generates item IDs. 
During this process, constrained beam search is applied over the vocabulary of item IDs, with complexity approximately proportional to
\(B \times L \times V_{item}\).
Here, \(B\) is the beam size, \(L\) is the length of the generated ID sequence (typically 1 for a single item ID), and \(V_{item}\) is the total number of items in the corpus.

\begin{table*}[t]
\setlength{\abovecaptionskip}{0.3cm}
\setlength{\belowcaptionskip}{0cm}
\caption{Overall performance of GenSR and other baselines in the full-ranking setting. Bold signifies the best performance among the baselines and GenSR, and underline represents the second-best method. * denotes statistically significant improvements of GenSR over the best baseline, according to t-tests with a significance level of $p$ < 0.01.}
\label{tab:rank}
\centering
\resizebox{\textwidth}{!}{ 
\begin{tabular}{ll|cccc|cccc}
\toprule
\multicolumn{2}{c|}{}                                                                & \multicolumn{4}{c|}{\textbf{Amazon}}                                                                                                                              & \multicolumn{4}{c}{\textbf{KuaiSAR}}                                                                                                                              \\
\multicolumn{2}{c|}{\multirow{-2}{*}{\textbf{Method}}}                               & \textbf{R@5}                      & \textbf{R@10}                     & \textbf{N@5}                        & \textbf{N@10}                       & \textbf{R@5}                      & \textbf{R@10}                     & \textbf{N@5}                        & \textbf{N@10}                       \\ \midrule

\multirow{3}{*}{\textbf{Recommendation}} 
& \textbf{TIGER~\cite{rajput2023recommender}}        & 0.0108 & 0.0173 & 0.0101 & 0.0142 & 0.0149 & 0.0213 & 0.0138 & 0.0201 \\
& \textbf{BSR~\cite{penha2024bridging}} & 0.0027 & 0.0037 & 0.0016 & 0.0019 & 0.0052 & 0.0081 & 0.0043 & 0.0071 \\
& \textbf{GenSR} & \textbf{0.0153*} & \textbf{0.0211*} & \textbf{0.0136*} & \textbf{0.0198*} & \textbf{0.0187*} & \textbf{0.0254*} & \textbf{0.0181*} & \textbf{0.0250*} \\ \midrule

\multirow{3}{*}{\textbf{Search}} 
& \textbf{GenRet~\cite{sun2023learning}}     & 0.0221 & 0.0311 & 0.0197 & 0.0277 & 0.0112 & 0.0181 & 0.0099 & 0.0173 \\
& \textbf{BSR~\cite{penha2024bridging}}  & 0.0072 & 0.0080 & 0.0058 & 0.0059 & 0.0074 & 0.0109 & 0.0066 & 0.0095 \\
& \textbf{GenSR} & \textbf{0.0268*} & \textbf{0.0362*} & \textbf{0.0223*} & \textbf{0.0305*} & \textbf{0.0181*} & \textbf{0.0259*} & \textbf{0.0144*} & \textbf{0.0216*} \\ 
\bottomrule
\end{tabular}

}
\end{table*}

We compare GenSR with three representative generative models under the full-ranking setting: 
\begin{itemize}
    \item \textbf{BSR}~\cite{penha2024bridging} is a generative joint S\&R model that outputs items as plain IDs, without leveraging content semantics.
    \item \textbf{TIGER}~\cite{rajput2023recommender} is a generative recommendation model that assigns each item a discrete Semantic ID and trains a sequence-to-sequence model to predict the next item’s Semantic ID in a user’s history
    \item \textbf{GenRet}~\cite{sun2023learning} is a generative retrieval model that learns to tokenize documents into short discrete identifiers and generate these identifiers for a given query.
\end{itemize} 

These baselines include both unified S\&R models and single-task generative approaches.
We evaluate all models on the Amazon and KuaiSAR datasets under the full-ranking setting, using the same top-$K$ metrics as reported in Section~\ref{sec:setting}.
The results are presented in Table~\ref{tab:rank}.
As expected, all models experience a performance drop compared to the main setting due to the significantly larger candidate space.
Nevertheless, GenSR consistently outperforms the generative baselines by a substantial margin across all metrics and datasets, owing to its superior dual representation learning and unified training strategy.
Overall, these results strongly highlight GenSR’s remarkable generalization ability across different evaluation scenarios. 
While the typical candidate re-ranking setting remains widely used in unified S\&R task, GenSR also demonstrates its capacity to perform effectively in the generative full-ranking setting. 
This underscores GenSR's effectiveness and its potential to serve as a robust generative paradigm in search and recommendation tasks.


\subsection{Mutual Information Analysis (RQ2)}
In this section, we evaluate mutual information from two complementary angles.
First, we provide quantitative evidence by estimating the mutual information gains directly.
Second, we offer a qualitative geometric view via parameter-space visualization, showing how GenSR’s subspaces disentangle task manifolds and better align inputs with task-specific outputs, thereby increasing the mutual information between them.

\subsubsection{Quantifying mutual information gains}
\label{sec:empirical_analysis}

\begin{table}[t!]
\small
\setlength{\abovecaptionskip}{0cm}
\setlength{\belowcaptionskip}{0cm}
\caption{Mutual Information Estimation.}
\label{tab:mutual}
\begin{tabular}{c|cc|cc}
\toprule
\multicolumn{1}{c|}{}                                                                & \multicolumn{2}{c|}{\textbf{Amazon}}                                                                                                                              & \multicolumn{2}{c}{\textbf{KuaiSAR}}                                                                                                                             \\
\multirow{-2}{*}{\textbf{Method}}                            & \textbf{Search}                      & \textbf{Rec.}                     & \textbf{Search}                        & \textbf{Rec.}                        \\ 
\midrule
\textbf{UniSAR}        &          0.9003                         &      0.8739                                         &   0.8514                              &               0.8475                  \\
\textbf{GenSR}                 &   \textbf{0.9378}                              &  \textbf{0.9217}                                &         \textbf{0.9251}                         &                 \textbf{0.9168}                 \\
\bottomrule
\end{tabular}
\end{table}             
To further support our theoretical claim directly and verify the conclusion in a broader complex real-world scenario,
we employ Mutual Information Neural Estimation (MINE) to empirically estimate the practical mutual information between the input $X$ and the task-specific outputs $Y_S$ and $Y_R$~\cite{belghazi2018mutual}. 
Our empirical findings are consistent with the theoretical conclusion.
Specifically, we use the Donsker-Varadhan representation of KL divergence to estimate the lower bound of mutual information as:
\begin{equation}
I(X; Y_t) \geq \mathbb{E}_{P(X, Y_t)}[f(x, y_t)] - \log \mathbb{E}_{P(X)P(Y_t)}[e^{f(x, y_t)}],
  \label{eq:MINE}
\end{equation}
where $f(x, y_t)$ is a neural network function trained to maximize this bound.

In our experiments, we calculate mutual information for both S\&R tasks under the discriminative and generative paradigms. 
For each task, we estimate $I(X; Y_t)$ by sampling data from the joint distribution $P(X, Y_t)$ and comparing it with samples from the independent marginal distributions $P(X)$ and $P(Y_t)$. 
This process provides a quantitative measure of how well the input and output distributions are aligned in each paradigm.
We compare GenSR with the most powerful discriminative paradigm baseline UniSAR in two public datasets: Amazon and KuaiSAR (see detailed dataset description and baseline description in Section~\ref{sec:setting}).
The results, presented in Table~\ref{tab:mutual}, show that the generative paradigm consistently achieves higher mutual information for both tasks. 
This improvement aligns with our theoretical analysis, which is attributed to the task-specific subspaces. 
We next provide a parameter-space visualization to complement these quantitative results.

\subsubsection{Parameter space visualization analysis}
\label{sec:visual}

We empirically analyze the model's parameter space to deeply investigate and visually support our core claim: 
the proposed generative paradigm can effectively enhance the mutual information between the model input 
$X$ and task-specific output $Y_t$. 
We present our analysis in three steps: 
(1) the inherent differences in the original data distributions of the search and recommendation tasks; 
(2) the optimization challenges posed by the discriminative paradigm under a shared parameter space, which may lead to low mutual information; 
and (3) how the generative paradigm naturally achieves high mutual information by partitioning the parameter space. 
All analyses are conducted based on the KuaiSAR dataset, as it is the only real-world, publicly available dataset, offering authentic complexity and user behavior patterns and avoiding biases that may be introduced by synthetic data, thereby making our observations more representative.

\begin{figure}
  \setlength{\abovecaptionskip}{0cm}
  \setlength{\belowcaptionskip}{0cm}
  \centering
  \includegraphics[width=\linewidth]{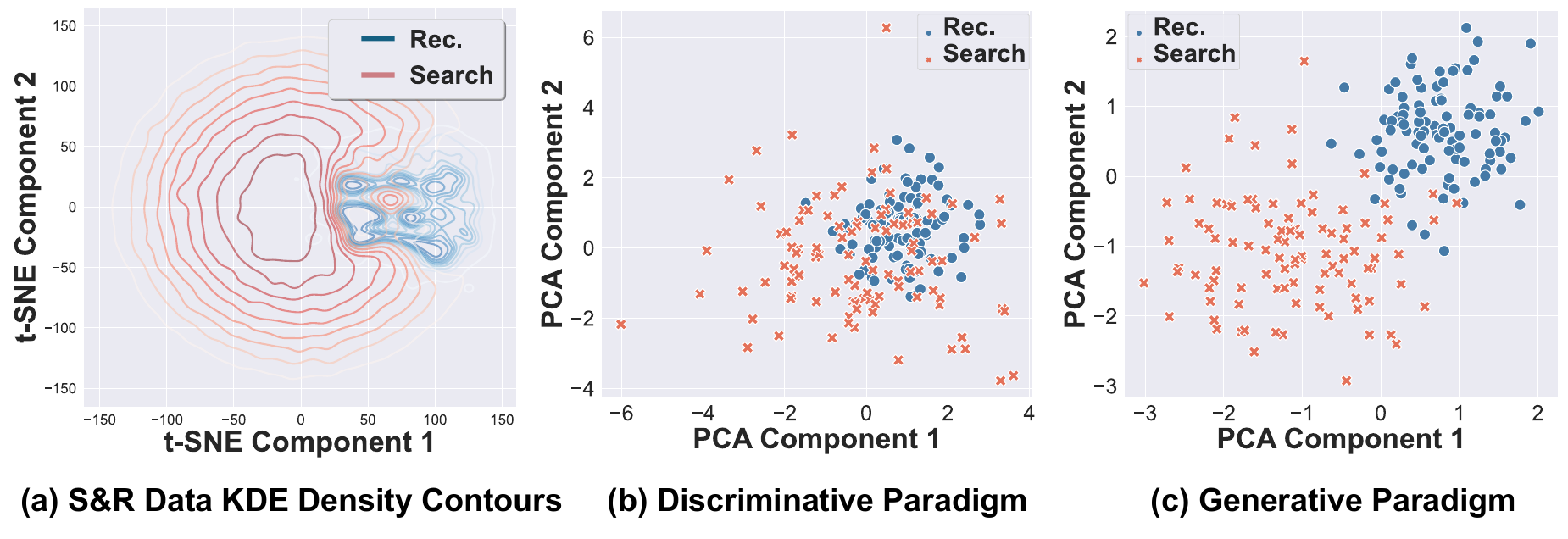}
  \caption{Parameter space visualization under KuaiSAR.}
  \label{fig:visual}
\end{figure}

To examine the raw data distributions of S\&R tasks, we utilize Kernel Density Estimation (KDE) to approximate $P(X_S, Y_S)$ and P$(X_R, Y_R)$~\cite{terrell1992variable}, providing a non-parametric representation of S\&R data distributions.
Specifically, we concatenate $X$ and $Y$ vectors for each sample, and apply KDE with Gaussian kernels to estimate the probability density. 
To visualize these distributions, we first use t-SNE to reduce the data into a two-dimensional space and then compute KDE in the resulting embedding space~\cite{van2008visualizing}. 
The KDE density contours for S\&R tasks are plotted in Figure~\ref{fig:visual}(a), revealing distinct density patterns that suggest differences in the underlying distributions of the two tasks.

Given the observed differences in task distributions, we further investigate the parameter space when unifying S\&R tasks under the discriminative paradigm.
Specifically, we extract the model’s hidden states for S\&R tasks and use Principal Component Analysis (PCA) to visualize the activation distribution in two dimensions as shown in Figure~\ref{fig:visual}(b).
The results indicate that the hidden state distributions of S\&R tasks under the discriminative paradigm exhibit significant overlap. 
This suggests that the model is forced to learn and represent two fundamentally different tasks within a single shared parameter space. 
Such enforced sharing leads to limitations in representational capacity: 
the model struggles to learn sufficiently distinct, task-specific representations for each task. 
From an information-theoretic perspective, this overlap causes the task-relevant information in the input $X$ to become entangled, thereby limiting the upper bound of mutual information. 
When the model cannot effectively differentiate or capture the distinct patterns of each task, it becomes difficult to find optimal gradient directions during training, which is the root cause of gradient conflicts (which will be quantified in Section~\ref{sec:gradient}).

Finally, we analyze the parameter space under our generative paradigm in Figure~\ref{fig:visual}(c).
The generative paradigm naturally partitions the parameter space into task-specific subspaces that align with the separate distributions of S\&R tasks. 
This separation yields multiple advantages:
(1) \textbf{Enhanced Mutual Information}:
The independence of subspaces ensures that input features are mapped more clearly and accurately to their corresponding task outputs. This allows each task’s representation learning and optimization to proceed without interference from the other, significantly enhancing the mutual information between input $X$ and task-specific output $Y_t$. 
This enriched mutual information translates directly into stronger task-specific representation capacity and improved prediction accuracy.
(2) \textbf{Mitigated Gradient Conflicts}:
By partitioning the space, the optimization gradients for the search and recommendation tasks can evolve independently within their own "dedicated" subspaces. 
This fundamentally reduces direct conflicts between task-specific optimization goals, enabling the model to converge more stably and efficiently while maximizing performance for both tasks.

\subsection{Gradient conflict analysis (RQ3)}
\label{sec:gradient}
In this section, we further empirically evaluate the degree of gradient conflict in both the discriminative paradigm and our proposed method to support that our method alleviates gradient conflict compared to the discriminative paradigm. 

To assess this effect, we compare our approach with the strongest discriminative baseline, UniSAR. 
At each training step, both models are supplied with the same mini-batch. 
We compute the loss for the search task and the recommendation task separately for the current mini-batch. 
Then, from both models (GenSR and UniSAR), we extract the corresponding gradients from the same self-attention layer in the transformer encoder. 
This layer is typically crucial for learning high-level representations and is where task-shared information is most concentrated, making it a likely site for gradient conflict.
And then, the extracted high-dimensional gradient tensors are flattened into one-dimensional vectors.
We compute the cosine similarity between the two flattened gradient vectors.
A higher cosine value indicates less conflict between the two tasks. 

\begin{figure}
  \setlength{\abovecaptionskip}{0cm}
  \setlength{\belowcaptionskip}{0cm}
  \centering
  \subfigure{
    \includegraphics[width=0.48\linewidth]{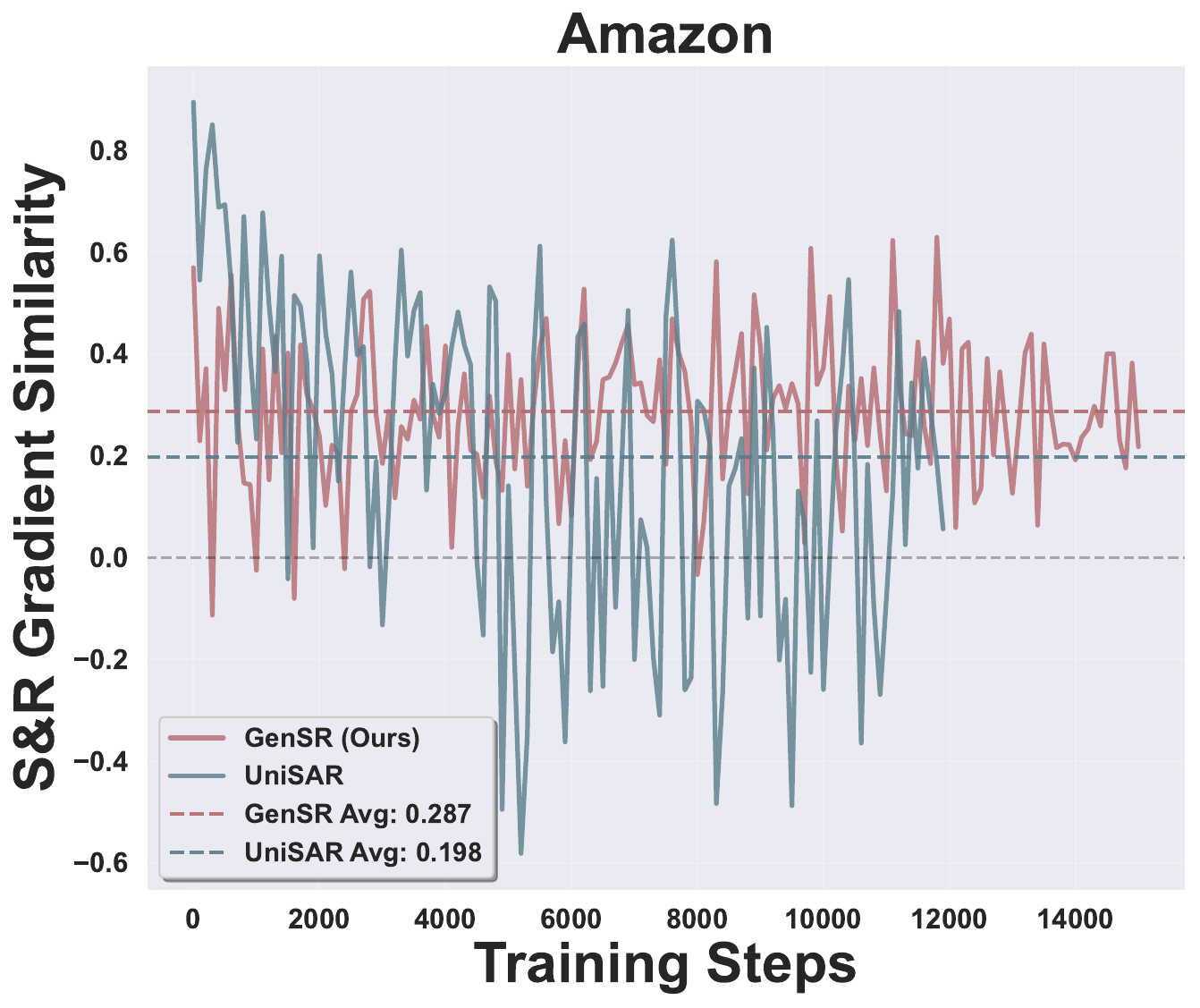}}
  \hfill
  \subfigure{
    \includegraphics[width=0.48\linewidth]{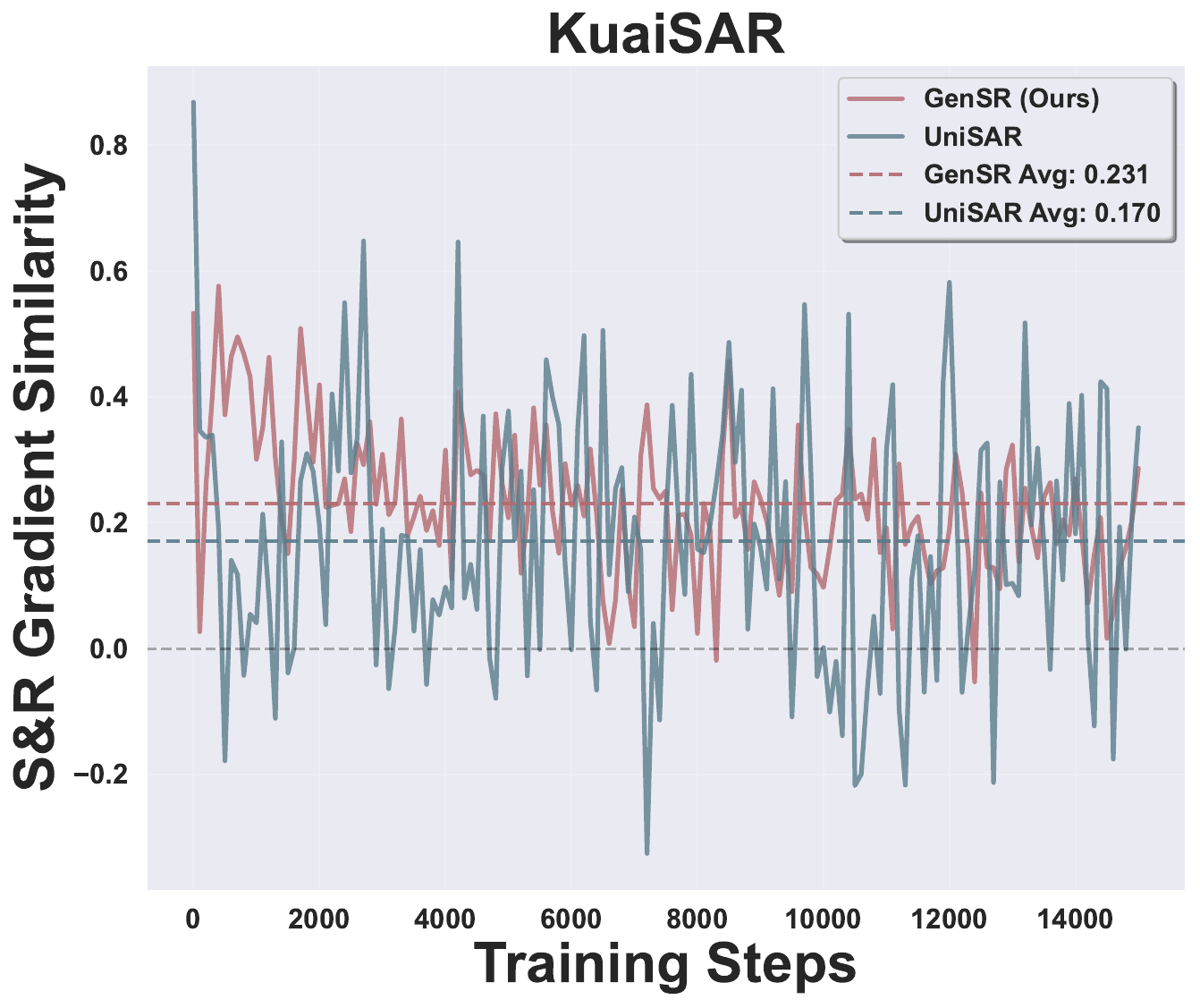}}
  \caption{Gradient Conflict Analysis.}
  \label{fig:conflict}
\end{figure}

Figure~\ref{fig:conflict} shows the similarity distribution across the training process. 
The key trends and conclusions are as follows:
\begin{enumerate}
    \item \textbf{Fluctuations and Negative Similarity in UniSAR}:
As a representative of the discriminative paradigm, UniSAR exhibits dramatic fluctuations in its gradient similarity curve, frequently dipping into the negative range.
These sharp fluctuations indicate frequent disagreements between the optimization objectives of S\&R tasks over shared parameters across training steps (mini-batches).
Additionally, the recurring dips into negative values (often around or below -0.5) clearly point to severe gradient conflicts. 
This suggests that at these steps, the gradient directions for search and recommendation are adversarial. 
Such opposing updates cause the model parameters to oscillate between conflicting objectives, slowing convergence, reducing training efficiency, and ultimately resulting in suboptimal performance. 
\item \textbf{Sustained Positive and Smooth Curve in GenSR}:
In contrast, GenSR maintains a consistently positive has a smaller amplitude in the gradient similarity curve throughout training.
The predominantly positive cosine similarities indicate that GenSR rarely encounters task conflicts; in most cases, the gradients of search and recommendation tasks are aligned or at least non-interfering. This means that even when the two tasks pursue different objectives, their updates to shared parameters remain largely compatible.
Additionally, the smoother evolution of the curve further reflects training stability and efficiency. In the absence of severe gradient opposition, model parameters converge more steadily toward globally optimal solutions.
\end{enumerate}

These contrasting trends clearly demonstrate that assigning search and recommendation to distinct subspaces significantly reduces gradient conflict and enables more stable optimization compared to the discriminative paradigm.


\begin{figure}
  \setlength{\abovecaptionskip}{0cm}
  \setlength{\belowcaptionskip}{0cm}
  \centering
  \subfigure{
    \includegraphics[width=0.48\linewidth]{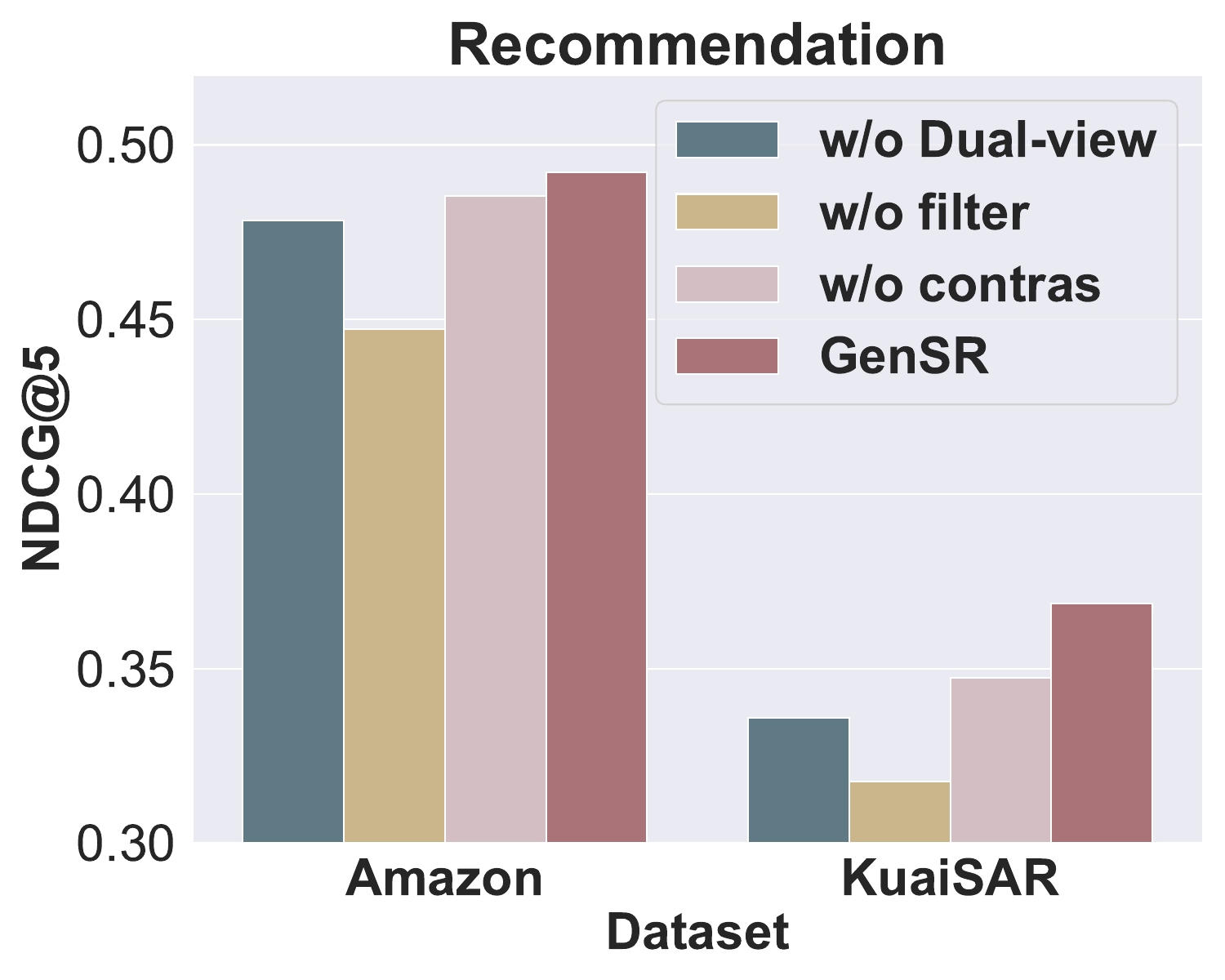}}
  \hfill
  \subfigure{
    \includegraphics[width=0.48\linewidth]{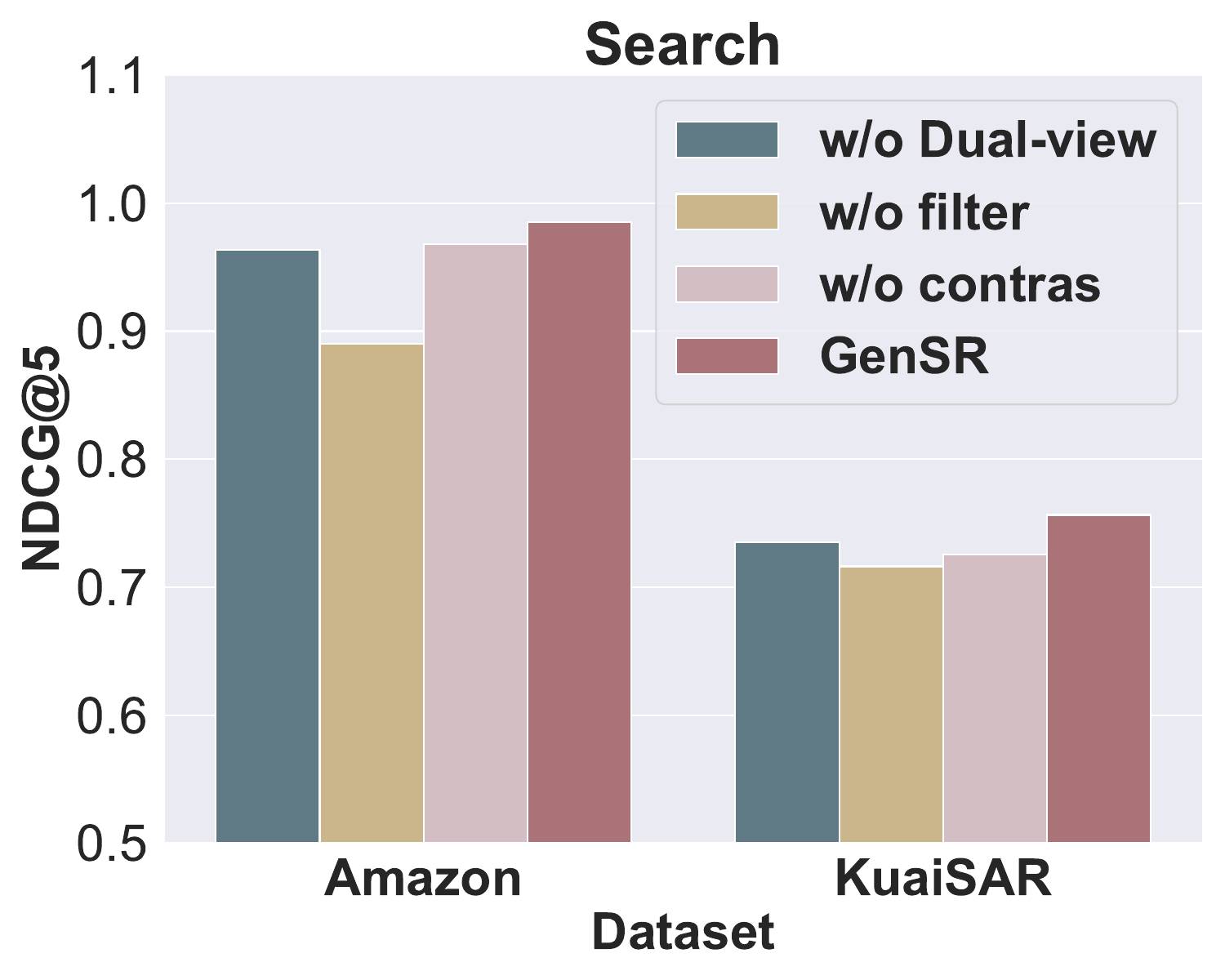}}
  \caption{Ablation study of different components in GenSR.}
  \label{fig:abl}
\end{figure}
\subsection{Ablation study (RQ4)}
\label{sec:ablation}
To systematically assess the individual contributions of each critical module within GenSR, we conducte a comprehensive ablation study. 
Specifically, our aim is to quantitatively analyze the independent impact of the dual-view representation learning, S\&R history soft filtering, and contrastive learning modules on the model's overall performance. 
Figure~\ref{fig:abl} presents the ablation experimental results, revealing the following key findings:
\begin{enumerate}
    \item \textbf{Contribution of Dual Representation Learning}: When the model is restricted to a single representation view (i.e., utilizing only the CF view or the semantic view), there is a substantial performance degradation across both recommendation and search tasks on both datasets. For instance, on the Amazon dataset, the NDCG@5 for the recommendation task dropped by approximately 26.5\% compared to the full model, while for the search task, NDCG@5 decreased by about 6.1\%. The KuaiSAR dataset exhibited similar performance drops.
    These findings underscore the intrinsic complementarity between CF and semantic information for unifying S\&R tasks. By independently learning representations from these two distinct views, GenSR effectively integrates the diverse information needed to support both S\&R tasks.  
    \item \textbf{Contribution of S\&R History Soft Filtering}:
   Excluding the S\&R history soft filtering module led to a noticeable decline in model performance. On the Amazon dataset, the NDCG@5 for the recommendation and search tasks decreased by approximately 12.2\% and 3.1\%, respectively. These results underscore the critical role of this filtering strategy in extracting relevant information from complex interaction histories. 
    \item \textbf{Contribution of Contrastive Learning}: The removal of the contrastive learning component resulted in a significant decline in model performance. On the Amazon dataset, the NDCG@5 for the recommendation and search tasks dropped by approximately 6.1\% and 1.0\%, respectively. These findings highlight the critical role of contrastive learning in aligning representations from different views. While dual-view representation learning independently captures CF and semantic information, contrastive learning acts as a crucial mechanism to integrate these complementary representations within a shared embedding space. 
    By encouraging the CF and semantic embeddings of the same user to be spatially closer while separating embeddings of different users, contrastive learning ensures both views jointly contribute to robust and comprehensive user preference modeling. This cross-view alignment is essential for enhancing the model’s generalization capability, as it prevents potential representation misalignment or disentanglement that may arise from independent learning. Ultimately, this mechanism maximizes the synergistic benefits of leveraging information from dual sources.
\end{enumerate}


\begin{figure}
  \setlength{\abovecaptionskip}{0cm}
  \setlength{\belowcaptionskip}{0cm}
  \centering
  \subfigure{
    \includegraphics[width=0.48\linewidth]{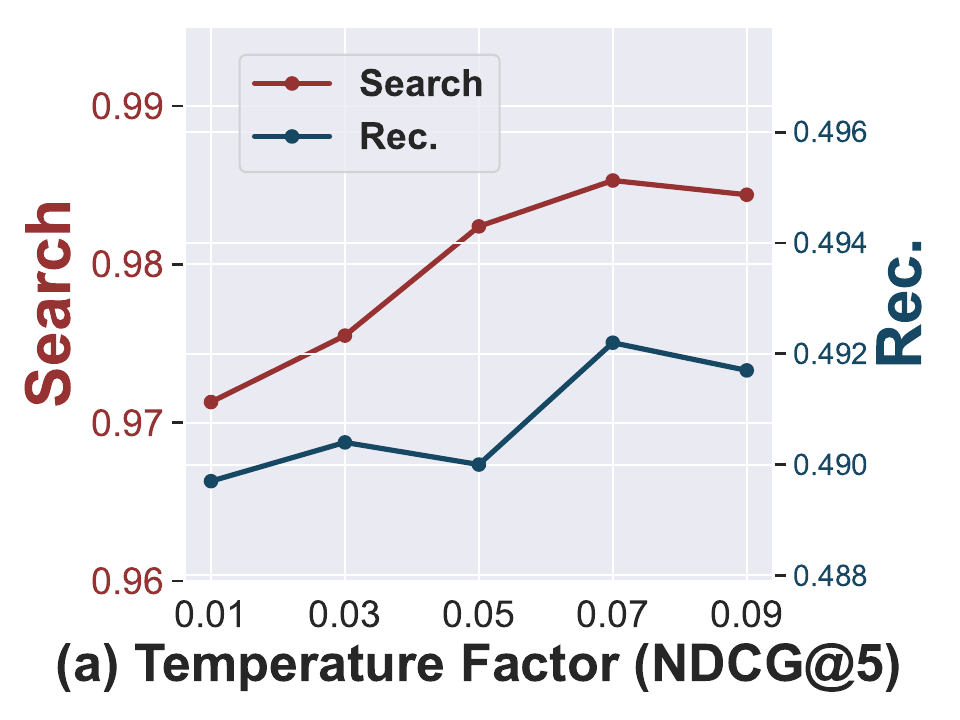}}
  \hfill
  \subfigure{
    \includegraphics[width=0.48\linewidth]{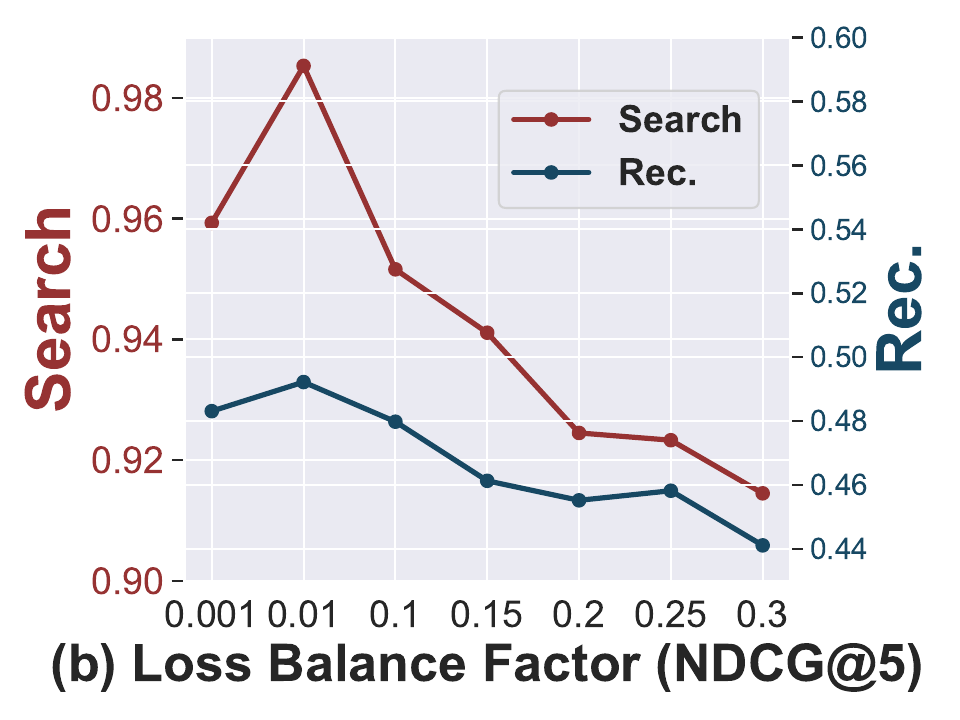}}
  \caption{Hyper-parameter analysis for the Amazon dataset.}
  \label{fig:hyper}
\end{figure}


\subsection{Hyper-parameter analysis (RQ4)}
\label{sec:hyper}
To thoroughly understand the robustness and optimal configuration of GenSR, we conducte a sensitivity analysis on two critical hyperparameters: the temperature factor (\(\tau\)) within the contrastive learning objective, and the loss balance factor (\(\beta\)) that governs the weighting of the contrastive loss relative to the main S\&R task loss. 
This analysis is performed on the Amazon dataset, with observed trends being consistent across the KuaiSAR dataset, thus the Amazon results are presented for conciseness in Figure~\ref{fig:hyper}. 
Our findings are as follows:
\begin{enumerate}
    \item The temperature factor \(\tau\) is a critical parameter in shaping the similarity distribution within the contrastive learning objective.  Specifically, it scales the logits before the softmax operation, thereby controlling the sharpness of the similarity distribution. A smaller \(\tau\) leads to a sharper distribution, assigning higher confidence to truly similar pairs and emphasizing fine-grained distinctions between positive and negative samples. Conversely, a larger \(\tau\) results in a smoother distribution, which softens the penalty for less similar negative pairs and can reduce the discriminative power. As shown in the Figure~\ref{fig:hyper}(a), GenSR's performance, measured by NDCG@5 for both recommendation and search tasks, exhibits a clear peak within a specific range of \(\tau\). 
    When \(\tau\) is too small (e.g., 0.01), performance significantly degrades (e.g., NDCG@5 for recommendation drops by approximately 1.5\% from peak). This is likely due to the contrastive loss becoming overly sensitive to minor differences or noise, potentially leading to feature collapse or focusing too intensely on easy negative samples, hindering the learning of generalized representations. Conversely, excessively large \(\tau\) values (e.g., 0.09) also lead to performance drops (e.g., NDCG@5 for recommendation drops by approximately 0.8\% from peak), as the loss function becomes less discriminative, failing to adequately distinguish between positive and negative pairs. This indicates that finding a balanced \(\tau\) is critical for effective contrastive learning. An appropriate \(\tau\) ensures that the model can sufficiently distinguish between positive (CF and semantic views of the same user) and negative samples (views of different users) without becoming overly sensitive or under-discriminating.
    \item The loss balance factor determines the relative importance of the contrastive learning loss \(L_c\) compared to the primary S\&R task generation loss \(L_{S\&R}\). It controls the trade-off between aligning the dual-view representations and optimizing for accurate task-specific outputs. Figure~\ref{fig:hyper}(b) illustrates the impact of \(\beta\) on performance. We observe that performance for both S\&R tasks generally improves as \(\beta\) increases from a very low value (e.g., 0.001) and then stabilizes or slightly declines after reaching an optimal range. 
    For instance, when \(\beta\) is too low (e.g., 0.001), the NDCG@5 for recommendation is significantly lower, suggesting that the contrastive loss has insufficient influence to properly align the CF and semantic views, thus failing to leverage their complementary strengths. As \(\beta\) increases into the optimal range, the model benefits from the improved aligned representations. However, excessively high \(\beta\) values can shift the model’s focus disproportionately toward embedding alignment, potentially marginalizing the primary S\&R task objectives. 
    This analysis underscores the importance of carefully selecting \(\beta\). A well-chosen value ensures the contrastive loss contributes sufficiently to robust representation learning and cross-view alignment without overshadowing the core S\&R objectives, maintaining a balance that supports both embedding alignment and task-specific accuracy.
\end{enumerate}


\section{Conclusion}
\label{sec:conclusion}

In this work, we have introduced GenSR, a novel generative paradigm that unifies search and recommendation by leveraging task-specific prompts to partition the model’s parameter space into specialized subspaces. 
This subspace partitioning framework encourages each task to learn distinct yet complementary representations, thereby enhancing mutual information and reducing gradient conflict within a unified model.
To enable effective subspace modeling, GenSR first constructs dual-view user representations that capture both collaborative and semantic signals across S\&R behaviors. 
These signals are aligned through contrastive learning and are fused with prompt-guided generation, allowing the model to adaptively generate task-specific outputs based on a unified interaction history. 
Unlike previous discriminative methods that rely heavily on manual architecture design or rigid parameter sharing, GenSR offers a more flexible and expressive alternative grounded in information-theoretic principles.

Extensive experiments on two public datasets, Amazon and KuaiSAR, demonstrate the effectiveness of our approach. 
GenSR consistently outperforms strong discriminative and generative baselines on both search and recommendation tasks. 
In addition to superior ranking performance, our analysis shows that GenSR achieves higher estimated mutual information and better gradient alignment, confirming that the proposed method improves the model representation capability and mitigate gradient conflict between two tasks.

While our framework is grounded in an information-theoretical analysis and employs task-specific subspaces to increase mutual information, there is potential to further optimize these subspaces from an optimization perspective. 
In future work, we aim to explore advanced optimization techniques to improve the alignment and mutual information between these task-specific subspaces.

\appendix
\section*{Appendix (Proof of Theorem 5.1)}

\label{appendix:proof_separability_mi}

This appendix provides a detailed proof of Theorem~\ref{the:mi_separability}, aiming to theoretically demonstrate how our proposed generative paradigm, GenSR, effectively enhances mutual information in unifying S\&R tasks. 

We consider two paradigms for unifying S\&R: the discriminative paradigm with parameter \(w_\theta\) and the generative paradigm with parameter \(w_\phi\). Let the true parameters be \(w_\theta^*\) and \(w_\phi^*\), and let their estimates from data be \(\hat{w}_\theta\) and \(\hat{w}_\phi\) respectively. 
The parameter approximation is obtained by minimizing their respective loss functions:\footnote{For binary generation tasks (e.g., ``yes'' vs.\ ``no''), 
    the negative log-likelihood in the generative setting 
    aligns with the binary cross-entropy loss of the discriminative setting, 
    allowing the same format of loss.}
\[
\hat{w}_\theta = \arg\min_{w_\theta} L(w_\theta), \quad \hat{w}_\phi = \arg\min_{w_\phi} L(w_\phi).
\]

In the binary classification setting, it has been proved that the discriminative parameter can be represented as the difference of the generative parameters across classes (i.e., \(w_\theta^* = w_{\phi,1}^* - w_{\phi,0}^*\), where $w_{\phi,0}^*$ and $w_{\phi,1}^*$ denote the true parameters for class 0 (``yes'') and class 1 (``no'') )~\cite{prasad2017separability}. 
Hence, comparing parameter approximation errors for the two paradigms is meaningful.
To do that, we expand the loss functions around the respective true parameters using Taylor's expansion. 
We define the approximation errors as follows:
$\Delta_D = \hat{w}_\theta - w_\theta^*$, $\Delta_G = \hat{w}_\phi - w_\phi^*$.
For each paradigm, we have:
\[
    \nabla L_n(w_\theta^*) + \nabla^2 L_n(w_\theta^*) \Delta_D + R_D(\Delta_D) = 0, 
\]
\[
    \nabla L_n(w_\phi^*) + \nabla^2 L_n(w_\phi^*) \Delta_G + R_G(\Delta_G) = 0, 
\]

where \(R_D(\Delta_D)\) and \(R_G(\Delta_G)\) are the corresponding higher-order remainder terms that capture the nonlinear and cross-dimensional interactions. 
Following~\cite{prasad2017separability}, we assume that the remainder terms satisfy a uniform bound in the local neighborhood: for any fixed small perturbation $\Delta$, then:
\[
\|R_G(\Delta)\|_1 \leq \frac{1}{\gamma_G}\|\Delta\|_1^{\alpha} \quad \text{and} \quad \|R_D(\Delta)\|_1 \leq \frac{1}{\gamma_D}\|\Delta\|_1^{\alpha}, \quad \alpha \ge 2,
\]
where \(\gamma_G\) and \(\gamma_D\) are the separability constants for the generative and discriminative paradigms respectively. Since it has been proven that the generative paradigm enjoys higher separability (i.e. \(\gamma_G > \gamma_D\))~\cite{prasad2017separability}, we have
$
\|R_G(\Delta)\|_1 < \|R_D(\Delta)\|_1$, for any fixed  $\Delta$.

Assuming that the perturbations are small enough for the linear part of the expansion to dominate, we can apply the Taylor expansions~\cite{song2011automatic}:
\[
\Delta_G \approx -\left[\nabla^2 L_n(w_\phi^*)\right]^{-1} R_G(\Delta_G), \quad \Delta_D \approx -\left[\nabla^2 L_n(w_\theta^*)\right]^{-1} R_D(\Delta_D).
\]
Furthermore, the generative paradigm, through its explicit partitioning of the parameter space via task-specific prompts, and its inherent higher separability, results in a Hessian matrix with better conditioning during optimization:\footnote{Although this uniform bound is stated for an arbitrary (fixed) $\Delta$, in practice the error analysis is applied to the respective perturbations $\Delta_D$ and $\Delta_G$ in their own local neighborhood. Since both $\Delta_D$ and $\Delta_G$ are assumed to be small (and we are only using the bound locally), it is reasonable to apply the same type of bound separately to each—even though they are different. The uniformity of the bound assures us that the same structure holds regardless of which (small) perturbation is under consideration.}
\[
\left\| \left[\nabla^2 L_n(w_\phi^*)\right]^{-1} \right\| \leq \left\| \left[\nabla^2 L_n(w_\theta^*)\right]^{-1} \right\|.
\]
Thus, combining these observations we obtain
\begin{align*}
    \|\Delta_G\| &= \left\| \left[\nabla^2 L_n(w_\phi^*)\right]^{-1} R_G(\Delta_G) \right\| \le \left\| \left[\nabla^2 L_n(w_\phi^*)\right]^{-1} \right\| \, \|R_G(\Delta_G)\|_1 \\
    &< \left\| \left[\nabla^2 L_n(w_\theta^*)\right]^{-1} \right\| \, \|R_D(\Delta_D)\|_1 \approx \|\Delta_D\|,
\end{align*}
where the last approximate equality holds when the remainder terms are small compared to the linear term. This chain of inequalities leads to the conclusion:
\begin{equation}
\label{eq:weight}
\|\hat{w}_\phi - w_\phi^*\| \leq \|\hat{w}_\theta - w_\theta^*\|.
\end{equation}
This indicates that the generative paradigm can achieve a lower parameter approximation error, meaning its estimated parameters are closer to the true parameters.

Consider a Gaussian model given by:
$Y = w^* X + b^* + \epsilon$, with $X \sim \mathcal{N}(\mu_X, \sigma_X^2)$, $\epsilon \sim \mathcal{N}(0, \sigma_\epsilon^2)$ independent of $X$,  
Building on prior work in information theory and statistical learning (e.g.,\cite{leung2006information,gabrie2018entropy}),
mutual information between input \(X\) and prediction \(\hat{Y}\), parameterized by model parameter \(w\), is expressed as:
\begin{equation}
    \label{eq:mutual_info}
    I_w(X; \hat{Y}) = \frac{1}{2} \ln\left(1 + \frac{w \sigma_X^2}{\sigma_\epsilon^2} \right).
\end{equation}

This formula illustrates that the model's representation capability, as reflected by mutual information, is related to the magnitude of the parameter \(w^*\) and the variances of the input and noise.
Specifically, when parameter estimates \(\hat{w}\) are closer to the true parameters \(w^*\), the model can better capture the information embedded in the data, leading to higher mutual information.
Therefore, tnder comparable noise variances~\cite{hastie2009elements}, 
as parameter estimates \(\hat{w}\) approach the true parameter \(w^*\), the mutual information \(I(X;\hat{Y})\) approaches the theoretical upper bound \(I(X;Y)\), which is identical for both paradigms.
Given that we have established Eq.(\ref{eq:weight}) and Eq.(\ref{eq:mutual_info}),
the generative paradigm achieves lower approximation error, leading directly to higher mutual information:

\[
    I_{w_\theta}(X; Y_S) + I_{w_\theta}(X; Y_R)
    \leq
    I_{w_\phi}(X; Y_S) + I_{w_\phi}(X; Y_R).
\]

{
\tiny
\bibliographystyle{ACM-Reference-Format}
\balance
\bibliography{bibfile}
}

\end{document}